\documentclass[a4paper,fleqn,usenatbib,,threeparttable]{mnras}

\usepackage[T1]{fontenc}
\usepackage{ae,aecompl}

\usepackage{graphicx}	
\usepackage{amsmath}	
\usepackage{amssymb}	
\usepackage{xspace}

\usepackage{natbib}
\usepackage[usenames]{color}
\usepackage{subfig} 
\usepackage{multirow}
\usepackage{array}
\usepackage{verbatim}
\usepackage{booktabs,caption,fixltx2e} 
\usepackage[flushleft]{threeparttable}
\usepackage{color}

\usepackage{amsmath}  
\let\oldAA\AA
\renewcommand{\AA}{\text{\normalfont\oldAA}}

\usepackage{dblfloatfix}

\title[Cosmic evolution of \mbh-\lhost]{H0LiCOW VII. Cosmic evolution of the correlation between black hole mass and host galaxy luminosity}

\author[X. Ding et al.]{\parbox{\textwidth}{
Xuheng~Ding,$^{1,2,3}$\thanks{E-mail: dxh@astro.ucla.edu}
Tommaso~Treu,$^{2}$
Sherry~H.~Suyu,$^{4,5,6}$
Kenneth~C.~Wong,$^{7}$
Takahiro~Morishita,$^{2,8,9}$
Daeseong~Park,$^{10}$
Dominique~Sluse,$^{11}$
Matthew~W.~Auger,$^{12}$
Adriano~Agnello,$^{2,13}$
Vardha~N.~Bennert,$^{14}$
Thomas~E.~Collett$^{15}$\\
}
\\
\parbox{\textwidth}{
$^{1}$School of Physics and Technology, Wuhan University, Wuhan 430072, China\\
$^{2}$Department of Physics and Astronomy, University of California,
Los Angeles, CA, 90095-1547, USA\\
$^{3}$Department of Astronomy, Beijing Normal University, Beijing
100875, China\\
$^{4}$Max-Planck-Institut f{\"u}r Astrophysik, Karl-Schwarzschild-Str.~1,
85748 Garching, Germany\\
$^{5}$Institute of Astronomy and Astrophysics, Academia Sinica, P.O.~Box 23-141, Taipei 10617, Taiwan\\
$^{6}$Physik-Department, Technische Universit\"at M\"unchen, James-Franck-Stra\ss{}e~1, 85748 Garching, Germany\\
$^{7}$National Astronomical Observatory of Japan, 2-21-1 Osawa,
Mitaka, Tokyo 181-8588, Japan\\
$^{8}$Astronomical Institute, Tohoku University, Aramaki, Aoba, Sendai 980-8578, Japan\\
$^{9}$Institute for International Advanced Research and Education, Tohoku University, Aramaki, Aoba, Sendai 980-8578, Japan\\
$^{10}$Korea Astronomy and Space Science Institute, Daejeon, 34055, Republic of Korea\\
$^{11}$STAR Institute, Quartier Agora - All\'ee du six Ao\^ut, 19c
B-4000 Li\`ege, Belgium\\
$^{12}$Institute of Astronomy, University of Cambridge, Madingley Road,
Cambridge CB3 0HA, UK\\
$^{13}$European Southern Observatories, Karl-Schwarzschild-Str.~2,
85748 Garching, Germany\\
$^{14}$Physics Department, California Polytechnic State University, San Luis Obispo CA 93407, USA; vbennert@calpoly.edu\\
$^{15}$Institute of Cosmology and Gravitation, University of Portsmouth, Burnaby Rd, Portsmouth PO1 3FX, UK
}}

\date{Accepted XXX. Received YYY; in original form ZZZ}

\pubyear{2017}

\begin{document}

\def\hequad{HE0435$-$1223}
\def\rxjquad{RXJ1131$-$1231}
\def\mbh{{$\mathcal M_{\rm BH}$}}
\def\efr{{$R_{\mathrm{eff}}$}}
\def\mstar{{$\mathcal M_*$}}
\def\lbulge{{$L_{\rm bulge}$}}
\def\lhost{{$L_{\rm host}$}}
\def\ltot{{$L_{\rm total}$}}
\def\glee{{\sc glee}}
\def\galfit{{\sc galfit}}
\def\lensfit{{\sc lensfit}}
\def\hst{\textit{HST}}
\def\lamLlam{$\lambda \rm L_{\lambda}$}
\newcommand{\kms}{km~s$^{\rm -1}$}
\newcommand{\ergs}{erg~s$^{\rm -1}$}
\newcommand{\Hb}{H$\beta$}
\newcommand{\Mgii}{Mg$_{\rm II}$}
\newcommand{\Civ}{C$_{\rm IV}$}

\label{firstpage}
\pagerange{\pageref{firstpage}--\pageref{lastpage}}
\maketitle

\begin{abstract}
  Strongly lensed active galactic nuclei (AGN) provide a unique
  opportunity to make progress in the study of the evolution of the
  correlation between the mass of supermassive black holes (\mbh) and
  their host galaxy luminosity (\lhost). We demonstrate the power of
  lensing by analyzing two systems for which state-of-the-art lens
  modelling techniques have been applied to deep \textit{Hubble Space
    Telescope} imaging data. We use i)~the reconstructed images to
  infer the total and bulge luminosity of the host and ii)~published
  broad-line spectroscopy to estimate \mbh\ using the so-called virial
  method. We then enlarge our sample with new calibration of
  previously published measurements to study the evolution of the
  correlation out to $z\sim4.5$. Consistent with previous work, we
  find that without taking into account passive luminosity evolution,
  the data points lie on the local relation. Once passive luminosity
  evolution is taken into account, we find that BHs in the more
  distant Universe reside in less luminous galaxies than
  today. Fitting this offset as \mbh/\lhost$\propto(1+z)^\gamma $, and
  taking into account selection effects, we obtain
  $\gamma = 0.6\pm0.1$ and $ 0.8\pm0.1$ for the case of \mbh-\lbulge\
  and \mbh-\ltot, respectively.  To test for systematic uncertainties
  and selection effects we also consider a reduced sample that is
  homogeneous in data quality. We find consistent results but with
  considerably larger uncertainty due to the more limited sample size
  and redshift coverage ($\gamma = 0.7\pm0.4$ and $0.2\pm0.5$ for
  \mbh-\lbulge\ and \mbh-\ltot, respectively), highlighting the need
  to gather more high-quality data for high-redshift lensed quasar
  hosts. Our result is consistent with a scenario where the growth of
  the black hole predates that of the host galaxy.
\end{abstract}




\section{Introduction}

It is commonly accepted that almost all the galaxies have a
supermassive black hole (BH) in their center, whose mass (\mbh) is
known to be correlated with the host properties.  The tight
correlations 
are usually, but not uniquely, explained as the results of their
co-evolution
\citep[e.g.,][]{Mag++98,F+M00,Geb++01b,M+H03,H+R04,Gul++09,Gra++2011,Beifi2012,Park15,Kormendy13}
\citep[see, however,][for a different view]{Peng:2007p22771,
  Jahnke:2010p6420}.  A powerful way to explore the origin of this
physical coupling and understand the role of active galactic nuclei
(AGN) feedback in galaxy formation is to measure the correlations
directly at high redshift and determine how and when they emerged and
evolved over cosmic time
\citep[e.g.,][]{TMB04,Sal++06,Woo++06,Jah++09,SS13,DeG++15}.

The most common technique used to estimate \mbh\ beyond the local
Universe ($z>0.1$) is the so-called virial method, based on the
properties of broad emission lines in type 1 AGN
\citep{Shen:2013p29308, Peterson2014SSRv}. However, the bright source
associated with the AGN makes the study of its host galaxy very
difficult. Strong gravitational lensing \citep[see, e.g.,][for
reviews]{CSS02,SKW06,Tre10,T+E15} stretches the host galaxy out from
the wings of the bright point source as point spread function (PSF),
providing a unique opportunity to infer its magnitude robustly
\citep{Pen++06qsob}.  However, in order to measure host luminosity
(\lhost) and construct the \mbh-\lhost\ correlation from strongly
lensed AGN, it is necessary to ensure that any systematic
uncertainties associated with the gravitational lens model can be
controlled to the desired level of accuracy.

Recently, \citet{H0licow6} studied the fidelity of the measurement of
lensed AGN host brightness through a set of extensive and realistic
simulations of {\textit {Hubble Space Telescope}} observation and lens
modeling. First, the mock images of the lensed AGNs in our sample (see
\citet{H0licow6}, Table 1) were generated as realistically as
possible. Second, the simulated AGN host galaxy images were
reconstructed with the state-of-the-art lens modelling tool (\glee
\footnote{Developed by \citet{S+H10} based on \citet{Suy++06} and
  \citet{Halk2008}.}). Third, by fitting the host magnitude with the
software \galfit\ \citep{Pen++02} and comparing the inference to the
input value, \citet{H0licow6} found that the \lhost\ can be recovered
with better accuracy and precision than the uncertainty on single
epoch \mbh\ estimates ($\sim 0.5$ dex) for hosts as faint as 2$-$4
magnitudes dimmer than the AGN itself.

In this paper, we apply our advanced techniques to two strongly lensed
systems (i.e. \hequad\ and \rxjquad), with excellent imaging data. The
host galaxy luminosity is inferred from the lens detailed model
developed as part of the H0LiCOW collaboration\footnote{{$H_0$ Lenses
    in COSMOGRAIL's Wellspring, \url{http://www.h0licow.org/}.}} with
the goal of measuring cosmological parameters from gravitational time
delays \citep{H0licow1, H0licow5}.  \mbh\ is inferred by applying a
set of self-consistent calibrations of the virial method to the broad
emission line properties measured by \citet{Sluse++2012}. In addition,
we combine our new measurements with a large sample of AGNs taken from
the literature and consistently recalibrated, and study the evolution
of the \mbh-\lhost\ relation for $146$ objects in the redshift range
$0<z<4.5$.  It is still unclear whether the bulge or the total
luminosity provides the tightest correlation with \mbh\
\citep{Jah++09, Bennert11, Park15}. Thus, we consider both of them in
this study.

This paper is organized as follows. We briefly describe the sample
selection in Section~\ref{sec:sample_select}. The host galaxy surface
photometry and the \mbh\ are inferred in Section \ref{sec:photometry}
and \ref{sec:bhmass}, respectively. In Section \ref{sec:result}, we
present our main result. Discussion and conclusion are presented in
Section \ref{sec:disc} and \ref{sec:sum}.
Throughout this paper, we adopt a standard concordance cosmology $H_0
= 70$ km s$^{-1}$ Mpc$^{-1}$, $\Omega{_m} = 0.30$, and $\Omega{_\Lambda} =
0.70$.  Magnitudes are given in the AB system.

\section{Sample selection}
\label{sec:sample_select}
First, we analyze the two quadruply-imaged AGN \hequad\ and
\rxjquad\ (hereafter HE0435 and RXJ1131) with source redshifts at
$1.693$ and $0.654$, respectively. Detailed information for these two
systems is given by \citet{H0licow1}.  Accurate lens models have been
derived in an effort to measure cosmological parameters from
gravitational time delays as described by \citet{H0licow4} and
\citet{Suy++13}. These models provide the reconstructions of AGN
hosts, from which in turn we estimate \lhost.

Second, we combine and compare our new measurements with those by
\citet{Pen++06qsob} (hereafter, P06). P06 explored the \mbh-\lhost\
based on 20 non-lensed AGNs and 31 gravitationally lensed AGNs
(including HE0435 and RXJ1131). P06 is so far the only paper in
  which the \mbh-\lhost\ relation has been comprehensively
  investigated using lensed AGNs observed with \hst. We note that
for the two systems in common, the \hst\ images used in our work are
much deeper than those used by P06, and the lens models are much more
detailed. Also, P06 was based on NIC2 images, as opposed to the much
more powerful more modern cameras used in our work. Therefore, our
measurements supercede those by P06 for these two
systems. Furthermore, we exclude MG~2016+112 because it is a type II
AGN \citep{Koo++02} and the black hole mass using the virial method
cannot be considered reliable. We also exclude the lens system
B2045+265 used by P06 because of the incorrect redshift identification
of the AGN spectrum by \citet{Fas++99b} (Nierenberg et al. 2017, in
preparation).

Third, we combine our new measurements with samples of non-lensed AGN
that have been measured by members of our team using the same
techniques as those applied here.  The samples consist of 52
intermediate redshift AGNs ($0.36< z< 0.57$) summarized by
\citet{Park15} (hereafter P15), 27 distant AGNs ($0.5< z<1.9$)
measured by \citet{Bennert11} and \citet{SS13} (hereafter, B11 and
SS13), and 19 local AGNs measurements \citep{Ben++10,
Peterson:2004p9469}. It is worth noticing that they are
so far the largest \hst\ imaging samples which are
carefully selected as moderate-luminosity AGN, for which the contrast
between nucleus and host galaxies is much more favorable for the
inference of \lhost\ than for high luminosity lensed quasars. Thus,
their host luminosities are measured with high accuracy even without
lensing.

Overall, our sample consists of two new lensed systems and
active galaxies from the literature, including elliptical and spiral
hosts with redshift up to 4.5.  This total sample of 146 objects is the largest compilation of
AGNs from \hst\ which are cross-calibrated to study the \mbh-\lhost\
relation.  The objects and their basic properties are listed in
Tab.~\ref{result} and \ref{local}.

\section{Surface photometry}
\label{sec:photometry}
In this section, we describe the measurement of host luminosity. For
HE0435 and RXJ1131, we first derived their host magnitude from the
reconstructed surface brightness maps in the source plane.  Then, we inferred
the rest-frame R-band luminosities based on their spectral energy
distribution (SED). For the other AGNs, we collected and homogenized
their luminosities from the literature.

\subsection{Surface photometry of HE0435 and RXJ1131} 
\label{sec:reconst}
We used the software \galfit\ to model the reconstructions from \citet{H0licow4} and
\citet{Suy++13}.
The reconstruction of HE0435 was fitted as the S\'ersic profile with
$n$ limited between $1-4$. It has been tested that this prior on $n$
does not bias the inference of magnitude \citep{H0licow6}. In the case
of RXJ1131 a clearly visible residual image was present and the
resulting parameters were physically acceptable when fitted with an
additional profile, we concluded that the host galaxy is composed of a
disk and a bulge. In this case, we fixed the reconstruction as
two-component S\'ersic profiles with $n$ equals to 1 and 4,
corresponding to exponential disk profile and \citet{deV48} profile,
respectively.  Although the luminosities of lens systems are corrected
from lensing magnification using a lens model, small differences
exist between models of different groups. The derived magnification
rarely differs by more than $20\%$. According to detailed simulations
presented by \citet{H0licow6}, the inferred values of \lhost\ can be
recovered with sufficient accuracy and precision to study the
\mbh-\lhost\ relation using our approach. Finally, we derived the
rest-frame R-band luminosity using a standard K-correction. These
steps are described below in more detail for each system.

\subsubsection{HE0435}
\label{ssec:0435}
HE0435 was imaged with \hst/WFC3-IR through filter F160W from program \hst-GO-12889 (PI: S.~H.~Suyu).
\citet{H0licow4} produced a set of twelve reconstructions for this
system, based on different assumptions, in order to estimate the
amplitude of systematic errors associated with these choices. In 9/12
cases the source plane resolution was set to 40$\times$40 pixels. For
the other three cases a higher resolution of 50$\times$50 pixels was
adopted. The reconstructions were based on an image plane size of
$\sim 1\farcs9$ square.

\begin{figure}
\centering
\subfloat[Lens reconstruction of HE0435 (left), the best-fit by \galfit\ (middle) and residual image (right).]{\includegraphics[trim = 0mm 20mm 0mm 0mm, clip,width=0.5\textwidth]{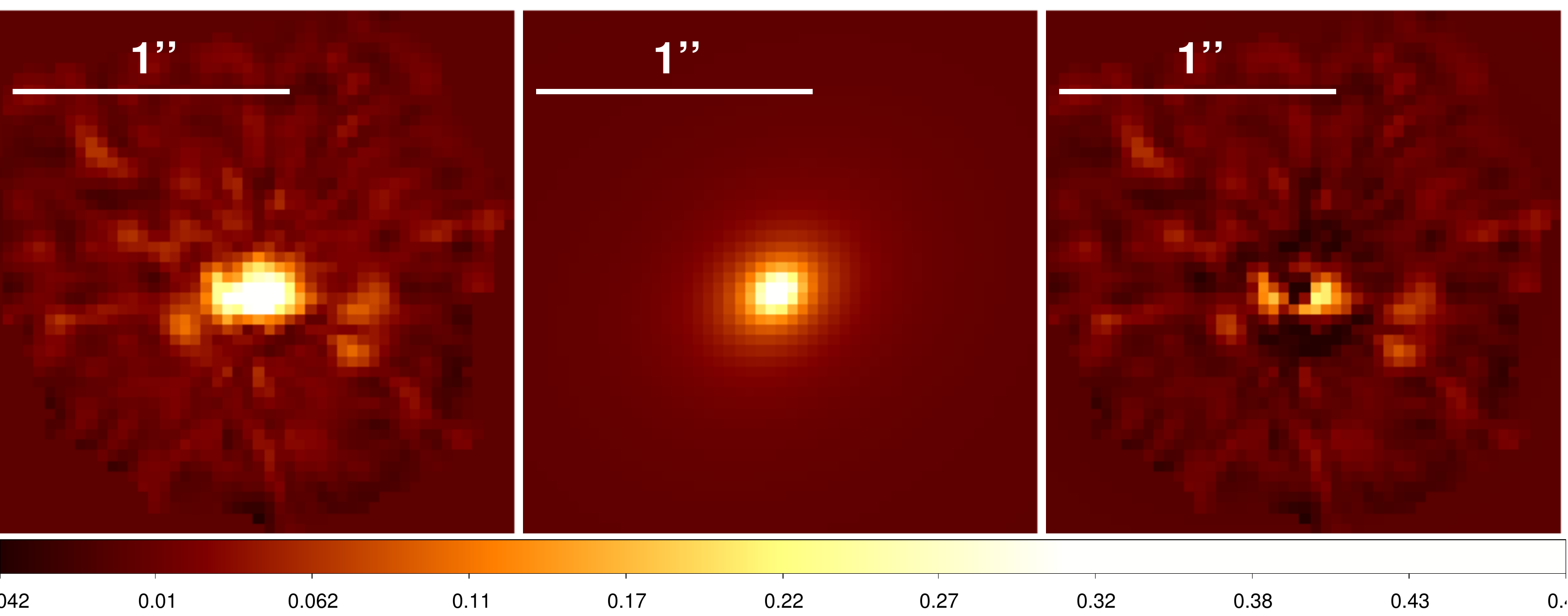}}\\
\subfloat[Observed image of HE0435 (left) and simulation with key parameters equal to inferred value (right).]{\includegraphics[trim = 0mm 20mm 0mm 0mm, clip,width=0.4\textwidth]{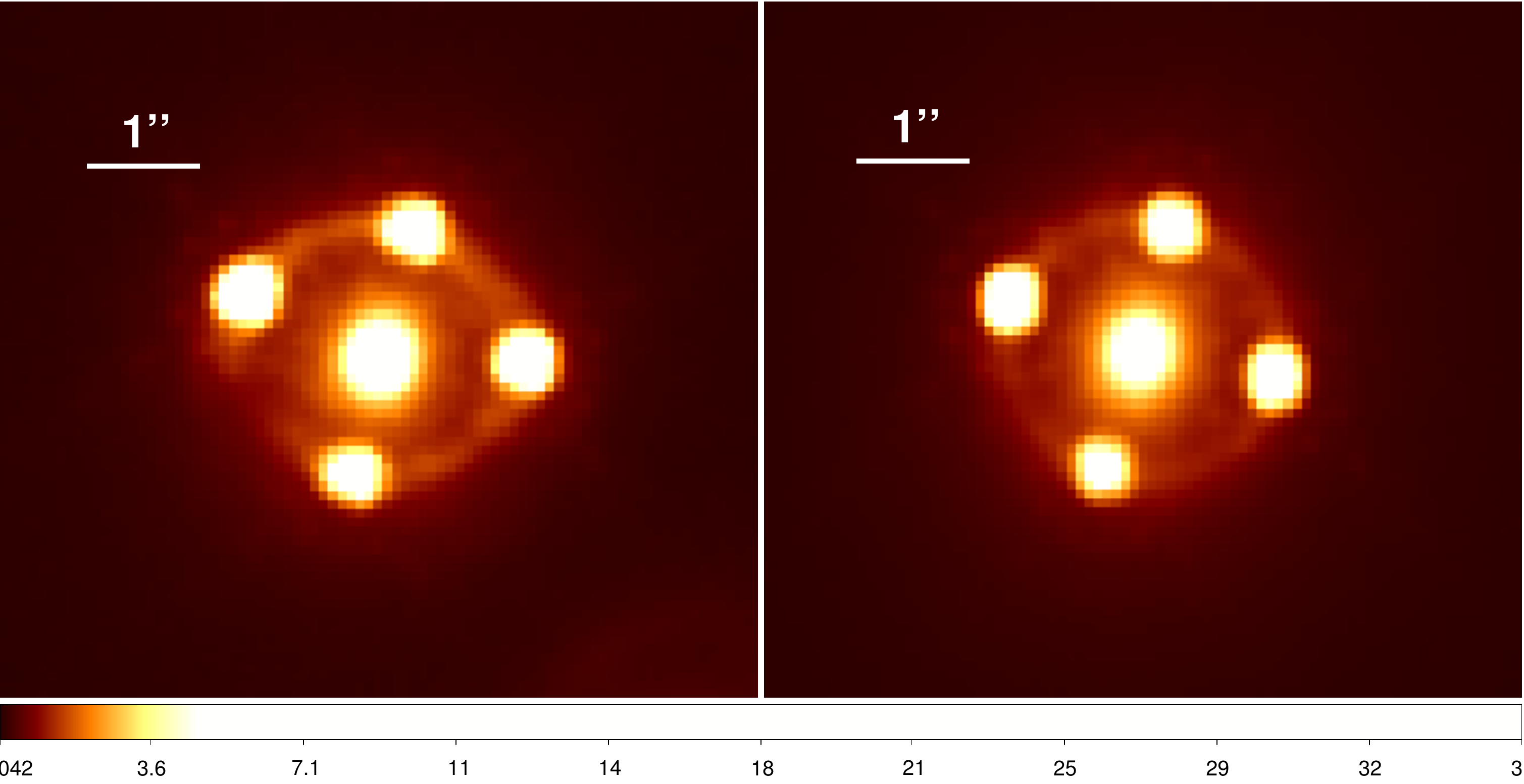}}
\caption{\label{fig:sim_0435} Illustration of the surface photometry study of HE0435,
presented with the same stretch for each panel, based on \hst/WFC3-IR
images through filter F160W.}
\end{figure}

By fitting each of the twelve reconstructions with single S\'ersic
profile, we summarized the inference and found the mean value and the
scatter of the host magnitudes are $m_{\rm host}=21.75 \pm0.13$; the
inferred effective radius and S\'ersic index are \efr$=0.82\pm0.14$
arcsecond; $n=3.94\pm0.14$, as shown in Tab.~\ref{infer}.
Furthermore, to test the type of the host galaxy, we fitted the
reconstructions as two-component S\'ersic profile. However, we
obtained unphysical results and no improvements in the fit indicating
that the host galaxy of HE0435 is consistent with being a pure
elliptical. One example of the reconstruction and its corresponding
\galfit\ best-fit are shown in Fig.~\ref{fig:sim_0435}, panel (a). We
also note that there is a small structure at the lower left of the
host. However, its brightness is negligible compared to the host which
do not affect the inference of the \lhost.  Interestingly, this could
correspond tidal features in the host galaxies.  If true, the mergers
could be related to triggered AGN activity. It is beyond the scope of
this work to pursue this further, but it would be intriguing to
simulate the hosts with merger signature and to see if they can be
recovered in the source reconstruction.

\begin{table}
\setcounter{table}{2}
\centering
    \caption{The inference of HE0435 and RXJ1131.}\label{infer}
     \begin{tabular}{ c c c c}
     \hline
     Object & magnitude& \efr & S\'ersic index ($n$) \\
     &&(arcsec)&\\
     \hline\hline
HE0435 & $21.75 \pm0.13$ & $0.82\pm0.14$ & $3.94\pm0.14$\\
RXJ1131$_{\rm disk}$& $20.07\pm0.06$ & 0.84$\pm0.09$ & fixed 1\\
RXJ1131$_{\rm bulge}$& $21.81\pm0.28$ & 0.20$\pm0.08$& fixed 4\\
     \hline
     \end{tabular}
\end{table}



We can verify the accuracy of our result by carrying out simulations
as described in our previous paper \citep{H0licow6}, using our
inferred parameters as input.
%
%
The observed and simulated HE0435 images are shown in
Fig.~\ref{fig:sim_0435}, panel (b). By repeating the analysis on the
simulated image, we recover the input value (input:
$m_{\rm host}=21.75$ mag; output 21.88 mag) showing an accuracy much
better than our target 1.25~mag (0.5~dex). We note that while in the
simulations the PSF is assumed to be perfectly known, for the real
data the PSF is inferred from the data using an iterative correction
procedure \citep[see][Suyu et al. in preparation]{Chen2016,H0licow4}.

Following P06, we made no corrections for dust extinction of the host
galaxy because they are likely to be small for a pure elliptical. The
observed magnitudes were then transformed to rest-frame R-band by
applying K-correction with Sbc template spectrum, using
\citet{Coleman1980} templates. We used the Sbc template because the
stellar populations cannot be older than a few Gyrs at this redshift
and the local elliptical template would be too red.  Nevertheless,
since the HE0435 is observed through the F160W filter, which roughly
corresponds to the rest-frame R-band at $z\sim1.5$, the K-correction
are only weakly dependent on the assumed SED (see Fig.~7 in P06), and
do not contribute significantly to the error budget. Finally, the best inferred
value of \lhost\ of HE0435 in rest-frame R-band is $\log$\lhost $=10.96$
which is very close to the one inferred in P06 (i.e. $\log$\lhost $=11.12$).

Lens models based on archival \hst\ Advanced Camera for Surveys (ACS)
in the filter F555W and F814W are also available from
\citet{H0licow4}. Unfortunately, due to the short exposure time, the
signal to noise ratio of the reconstructed host images in these bands
is insufficient to infer the luminosity robustly in these bands and
study the colors of the host. Thus they are not considered in this
study.

\subsubsection{RXJ1131}
\label{ssec:1131}
RXJ1131 is imaged with \hst/ACS through filter F814W. A set of seven
source resolutions including 50$\times$50, 52$\times$52, 54$\times$54,
56$\times$56, 58$\times$58, 60$\times$60, and 64$\times$64 pixels were
selected when modelling the host image into source plane
\citep{Suy++13}, with a frame size of $\sim 2\farcs9$ square\footnote{\citet{Suy++14}
  updated the model of RXJ1131.  Given the
  similarity in the composite and power-law model by \citet{Suy++14},
  a similar time delay distance is obtained (within $\sim2\%$, and
  hence spatial scaling of the source due to mass-sheet
  degeneracy). This means the inference of total flux of the host
  should be unchanged to within $\sim4\%$.}.

\begin{figure}
  \centering \subfloat[Source plane reconstruction of RXJ1131 (left),
  the best-fit by \galfit\ using a two-S\'ersic components
  profile (middle) and residual image (right).]
  {\includegraphics[trim = 0mm 20mm 0mm 0mm, clip, width=0.5\textwidth]{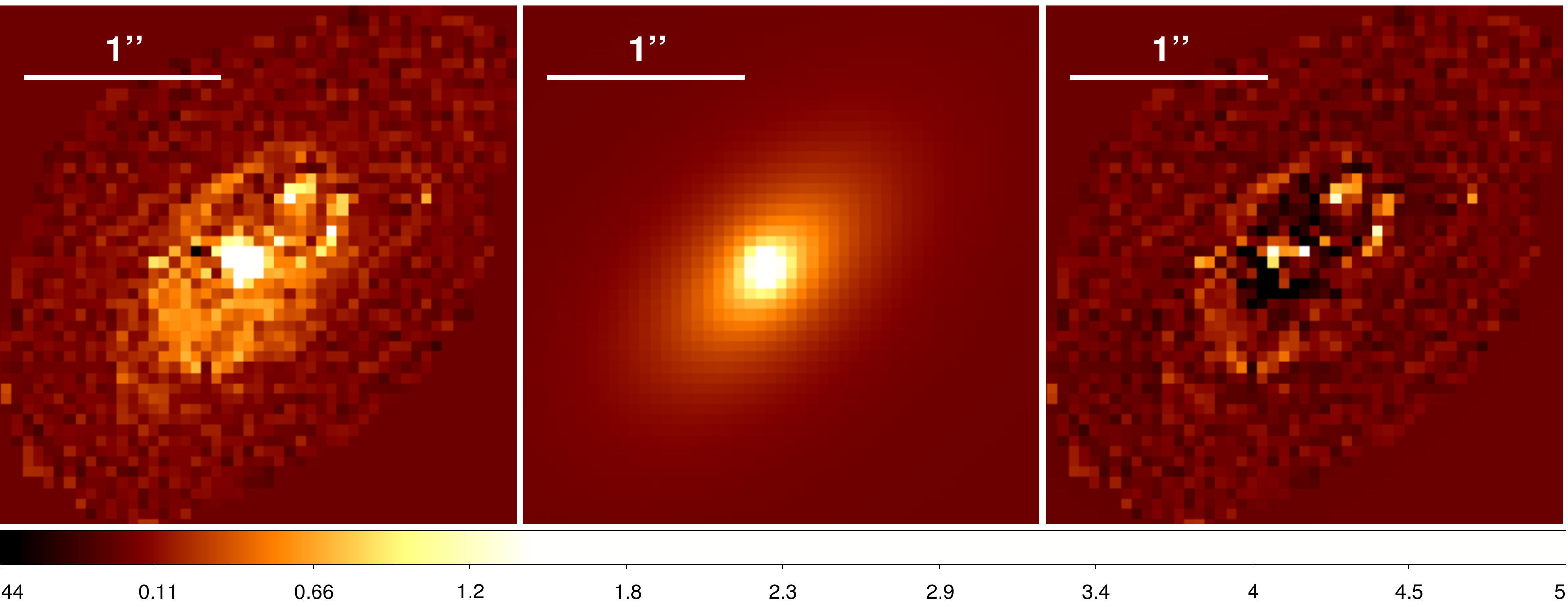}}\\
  \subfloat[Observed image of RXJ1131 (left) and simulated images with
  key parameters equal to inferred value (right).]
  {\includegraphics[trim = 0mm 20mm 0mm 0mm, clip, width=0.4\textwidth]{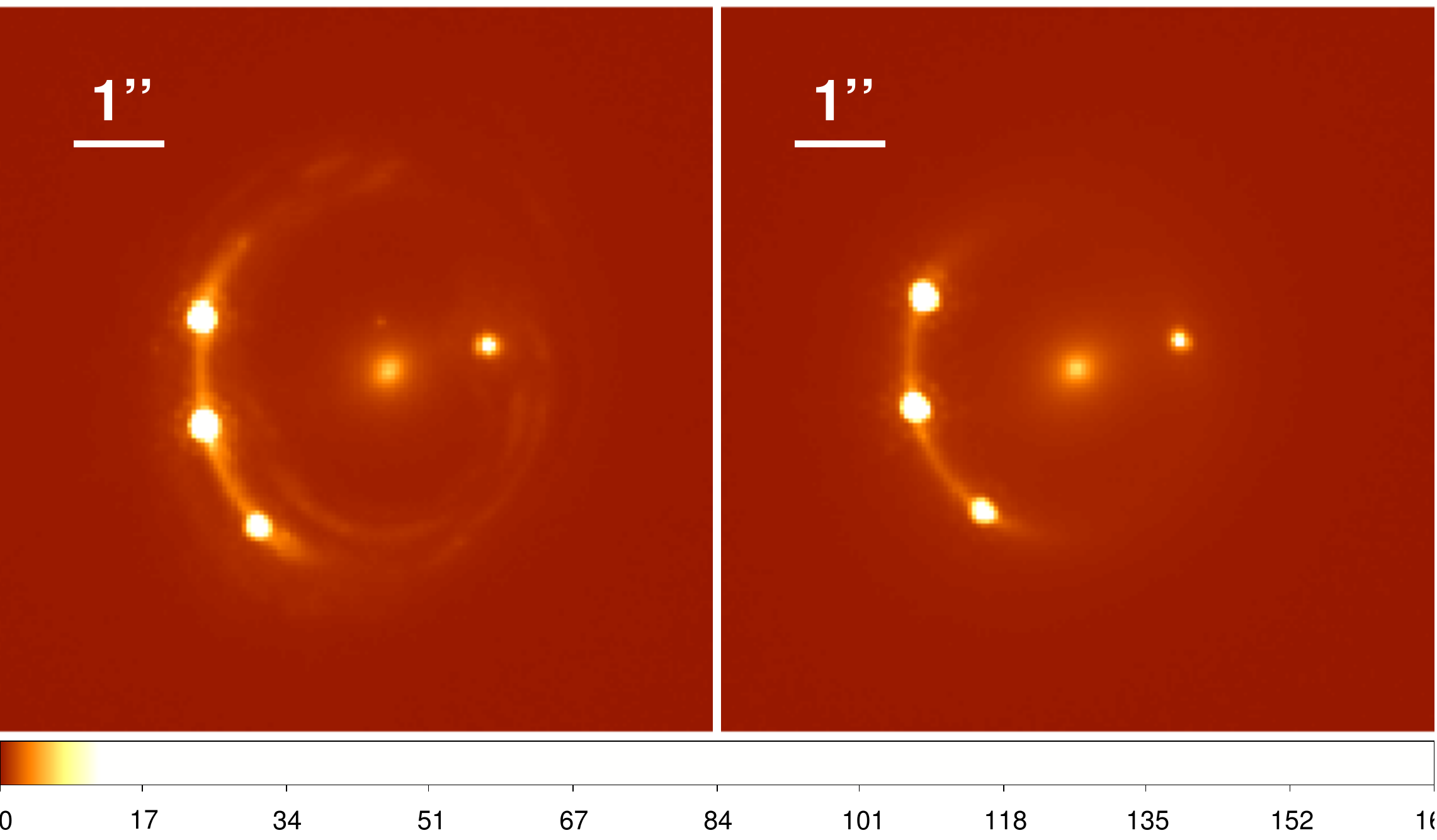}}\\
  \subfloat[Residual map of modelling the simulated reconstruction
  with single (left) and two-component (right) profile.]
  {\includegraphics[trim = 0mm 20mm 0mm 0mm, clip,
    width=0.35\textwidth]{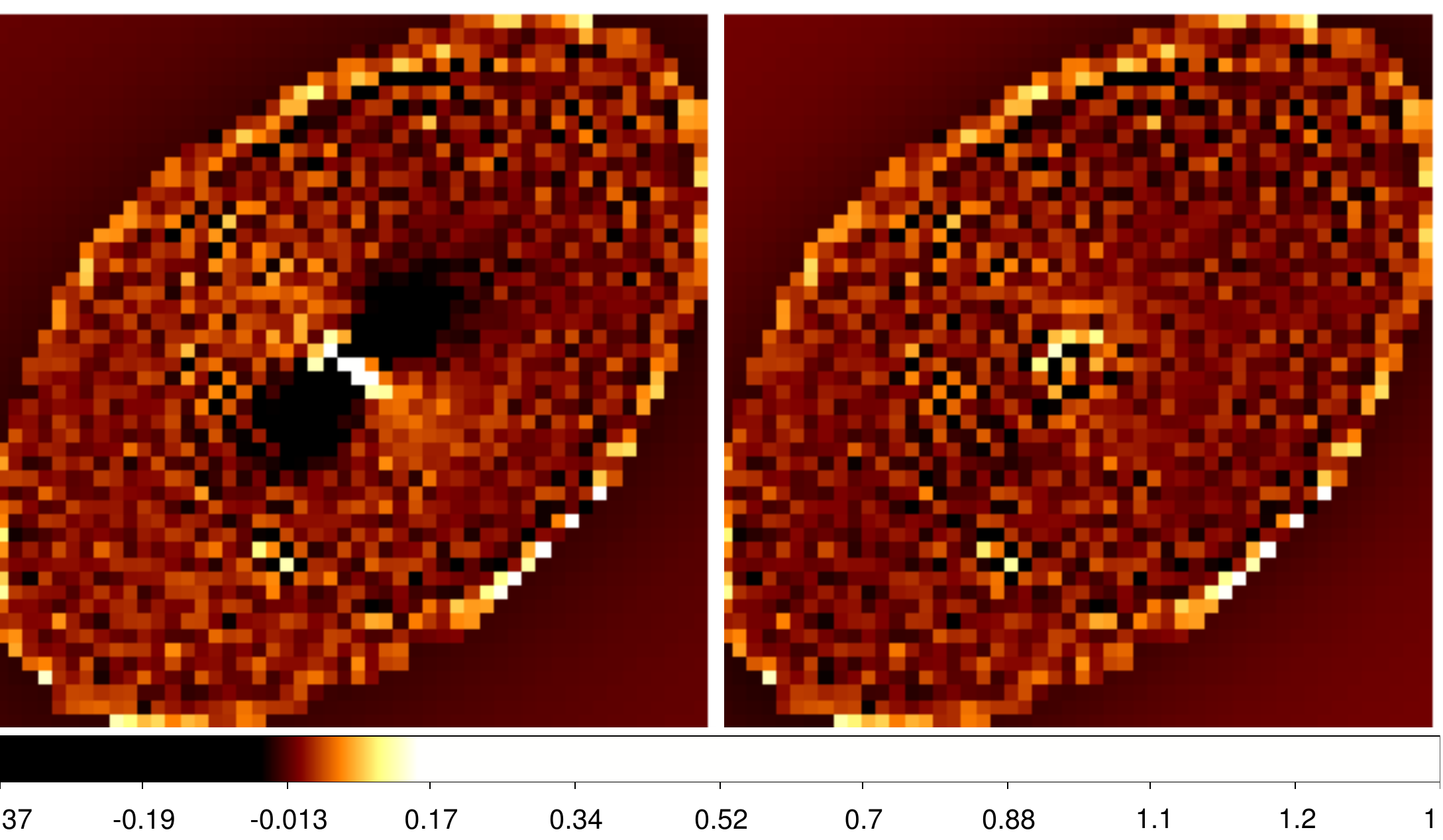}}
\caption{\label{fig:sim_1131} Illustration of the surface photometry study of RXJ1131,
presented with the same stretch for each panel, based on \hst/ACS images through filter F814W.}
\end{figure}

As noted by \cite{Suy++13}, all the reconstruction of the host show a
compact peak near the center (see Fig.~\ref{fig:sim_1131}, panel (a),
left panel), exhibiting the boundary line between the dominated area
of bulge and disk which indicates the host galaxy is a spiral
galaxy. Similarly, \citet{Claeskens+2006} reconstructed the host of
RXJ1131 and found it to be a spiral, disk dominated galaxy. Thus,
we fitted the reconstructions as two-component S\'ersic profiles, and
the inferred properties of the disk are $m_{\rm disk}=20.07\pm0.06$
mag; \efr$\__{\rm disk}=0.84\pm0.09$ arcsecond and the properties of
the bulge are $m_{\rm bulge}=21.81\pm0.28$ mag;
\efr$\__{\rm bulge}=0.20\pm0.08$ arcsecond, as summarized in
Tab.~\ref{infer}.  An example of the reconstruction and the best-fit
image are shown in Fig.~\ref{fig:sim_1131}, panel (a).

In the simulations of \citet{H0licow6}, the host of RXJ1131 was
assumed to be a single S\'ersic profile with the magnitude between
19.0 and 20.5. In this work, we simulate a more realistic
two-component profile, with key parameters (i.e.  $m_{\rm host}$ and
\efr) equal to the inferred values. The real and mock RXJ1131 image
are shown in Fig.~\ref{fig:sim_1131}, panel (b).  We first use a
single S\'ersic profile to fit the reconstruction, but applying this
model is a poor representation with an obvious residual in the central
image (i.e. Fig.~\ref{fig:sim_1131}, panel (c), left). This result
suggests the lens model of RXJ1131 reconstructs the host with
sufficiently high resolution to distinguish a bulge+disk model from a
single component.  Fitting with two-component S\'ersic profile, we
find that the residual map is much improved and both components can be
reconstructed accurately with our data and analysis techniques: input
$m_{\rm disk}=20.07$ mag and $m_{\rm bulge}=21.80$ mag; inferred
values are $m_{\rm disk}=20.37$ mag and $m_{\rm bulge}=22.07$ mag.

 
As for HE0435, we derived the rest-frame R-band magnitude using a
standard K-correction.  At the redshift of RXJ1131, the conversion to
R-band magnitude depends significantly on the adopted SED.  Therefore,
we determined the K-correction directly from the color of lensed host
arc, based on the multi-band SED fitting available in the archive (GO-9744;
PI: C.~S.~Kochanek).  The final estimations are
$\Delta {\rm mag}_{\rm disk}$(R$-$F814W)$\approx-0.3$ and
$\Delta {\rm mag}_{\rm bulge}$(R$-$F814W)$\approx-0.7$.  For detail,
see Appendix \ref{App1}. 


\subsection{Surface photometry for the literature samples}
In this section we describe our inference of the rest-frame R-band
luminosity for the P06 and P15 samples.

P06 used the \galfit\ (for non-lensed source) and \lensfit
\footnote{\lensfit\ is a version of \galfit\ that has been extended to
  fit lensed host galaxies while optimizing the mass model for the
  lens galaxy. For details, see P06.}  (for lensed source) softwares
to infer the brightness of the AGN hosts, describing the host galaxy
as single S\'ersic profile. 
In P06, they reported a single value of luminosity for each object,
suggesting
that the host galaxies are ellipticals. However, in our analysis, we
find the RXJ1131 is a spiral galaxy which suggests the approach in P06
may not be accurate for all the host galaxies. We will return to this
issue in Section \ref{sec:disc}.  Their measurements of absolute
magnitude are presented by P06 (Tab 3 and 4 therein) in rest-frame
R-band, Vega system. Thus, we transfer to AB system using
$m_{\rm AB, R} - m_{\rm Vega, R}=0.21$ \citep{Blanton07}.

Similarly, for the P15 sample, which includes the samples from B11 and
SS13, the host galaxy was fitted as a $n=4$ profile to model the bulge
component; an exponential disk profile was added if deemed
necessary. The rest-frame V-band luminosity is derived (see P15,
Tab. 4, Column 3) by applying the K-correction with an early-type
galaxy template spectrum. The same template is taken, we converted
their results to rest-frame R-band. As the scatter in V-R colors is
small, the associated uncertainty is estimated to be 0.16~mag
(i.e. 0.06~dex in luminosity). Likewise, the luminosities for 19 local
active galaxies are converted to rest-frame R-band.

Having obtained the R-band mag, the luminosity is derived by $\log
L_R/L_{R, \odot} = 0.4 (M_{R, \odot}-M_R)$, where $M_{R, \odot}=4.61$
\citep{Blanton07}. We summarized the homogenized R-band luminosities
in Tab.~\ref{result} and \ref{local}.

\section{Black hole mass} 
\label{sec:bhmass}
Assuming that the dynamics of the broad-line region (BLR) is dominated
by the gravity of the central supermassive black hole, \mbh\ can be
derived by applying the so-called virial method, based on the size of
the BLR ($R_{BLR}$) and the line-of-sight velocity width ($\Delta V$)
which can be inferred in turn from continuum luminosity and emission
line width, respectively. Usually, the \Civ($\lambda 1549$),
\Mgii($\lambda 2798$) and \Hb($\lambda 4861$) emission lines width and
their local continuum luminosities \lamLlam$(1300\AA)$,
\lamLlam$(3000\AA)$ and \lamLlam$(5100\AA)$ are used, respectively.

\citet{Sluse++2012}, P06, and P15 used different lines and different
calibrations of the virial method. Thus we need to cross-calibrate
them in order to avoid any systematic bias between the samples.



We first choose the recipe of P15 as the baseline:
\begin{eqnarray}
\label{eq:recipe_P15}
\log \left(\frac{\mathcal M_{\rm BH} }{M_\odot}\right)&~=~& 7.536 + 0.519 ~ \log\left(\frac{\lambda L_{5100}}{10^{44}~\rm erg~s^{-1}}\right) \nonumber\\
&~+~& 2 ~ \log \left(\frac{\sigma_{\textrm{\Hb}}}{1000~ \rm km~s^{-1}}\right) .
\end{eqnarray}
Then we align the self-consistent recipes (including emission lines
using \Hb\ and \Mgii) from \citet{McG++08} with this baseline, by
adding a small constant to the intercept (i.e. $-0.144$). In order to
cross-calibrate the \Civ-based estimator, we exploit the nine AGNs in
our sample for which both \Mgii\ and \Civ\ are available. We take the
\Civ\ recipe from P06 and add a small constant intercept
(i.e. $-0.331$) to match on average the value inferred from \Mgii.
Overall, we adopt the following virial formalism:
\begin{eqnarray}
\label{recipe}
\log \left(\frac{\mathcal M_{\rm BH}}{M_{\odot}}\right)&~=~& a+b \log \left(\frac{ \lambda \rm L _{\lambda_{line}}}{10^{44}{\rm erg~s^{-1}}}\right) \nonumber\\
&~+~& 2 \log \left(\frac{\rm FWHM(line)}{1000 ~{\rm km~s^{-1}}}\right) , 
\end {eqnarray}
with a\{\Civ, \Mgii, \Hb\}=\{6.322, 6.623, 6.882\},
b\{\Civ, \Mgii, \Hb\}=\{0.53, 0.47, 0.518\},
$\lambda_{line}$\{\Civ, \Mgii, \Hb\}=\{1350, 3000, 5100\}.
Having achieved a consistent cross-calibration, the \mbh\ is estimated
by adopting the emission line properties measured by
\citet{Sluse++2012}, P06 and P15.

For the 19 local AGNs, rather than using continuum luminosity,
$R_{BLR}$ was derived from time lags between continuum and
emission-line variations \citep{Peterson:2004p9469}. Thus, same as
P15, we adopt the reverberation-mapping \mbh\ measurements
with virial factor \citep[$\log f=0.71$,][]{Park++12, Bentz+2009}, noting that
they are the anchor for the virial method and thus are
inherently self-consistent.

\mbh\ estimates are listed in Tab.~\ref{result} and \ref{local},
together with details on the emission line used. For RXJ1131, since
the estimated \mbh\ using \Mgii\ and \Hb\ are very similar, we adopt
their average.  Moreover, we note that the values of \mbh\ for HE0435
and RXJ1131 inferred in this paper are larger than the estimates by
P06 ($\Delta \log $\mbh\ $= 0.25$ and $0.44$ for HE0435 and RXJ1131, respectively),
due to the fact that the properties of the emission lines of
these two systems have been revised upwards by \citet{Sluse++2012}
based on data of superior quality.

\section{Results}
\label{sec:result}
Following P15 and \citet{H0licow6}, for the distant objects, we adopt total
uncertainty for \lhost\ and \mbh\ of 0.2 dex ($\sim$0.5 mag) and 0.4 dex, respectively.

\subsection{The observed \mbh-\lhost\ relation}
\label{obs_ml}
\begin{figure*}
\centering
\begin{tabular}{ll}
 \hspace{-5.5em}
{\includegraphics[trim = 0mm 0mm 90mm 0mm, clip, height=0.6\textwidth]{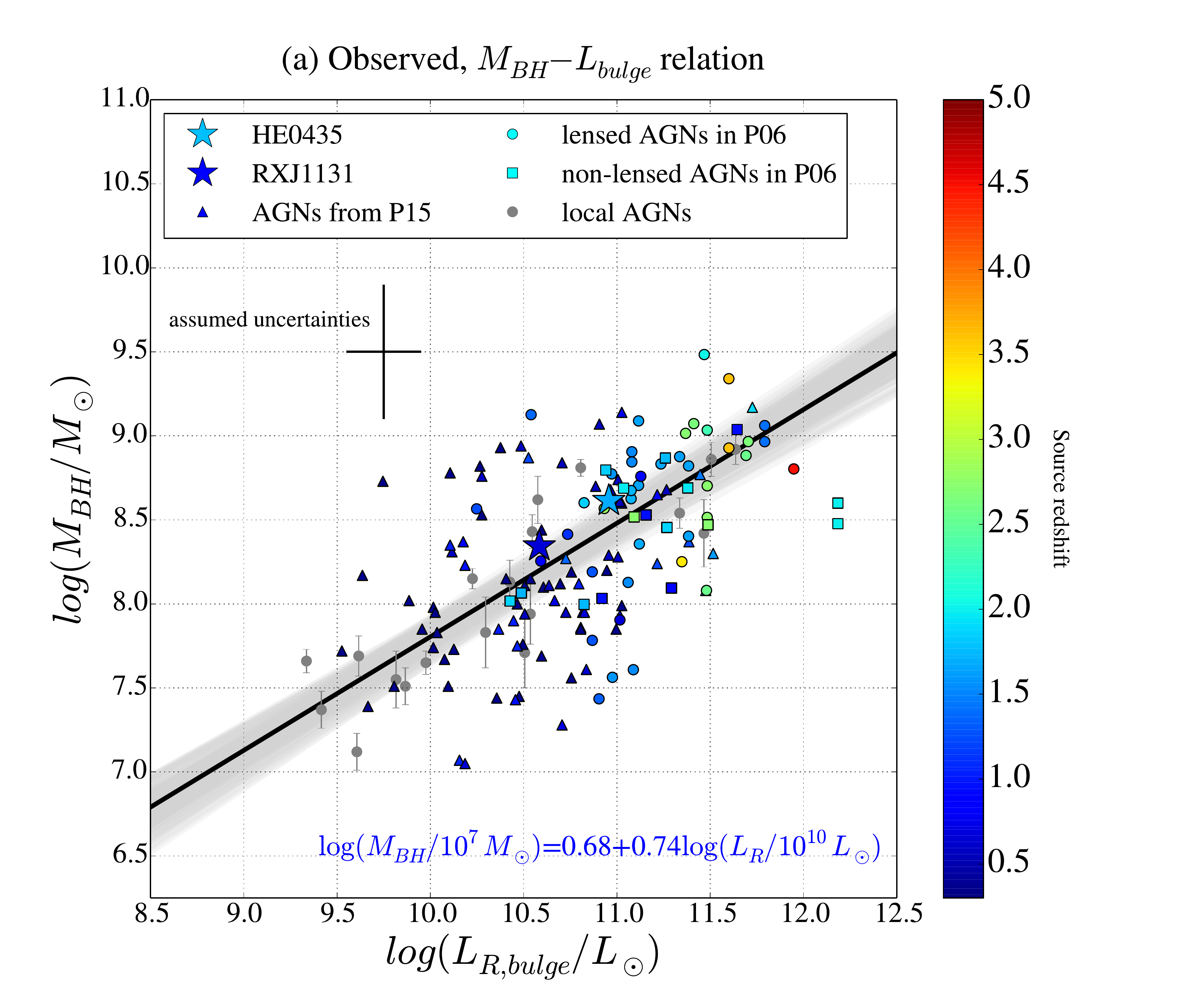}}&
{\includegraphics[trim = 46mm 0mm 0mm 0mm, clip, height=0.6\textwidth]{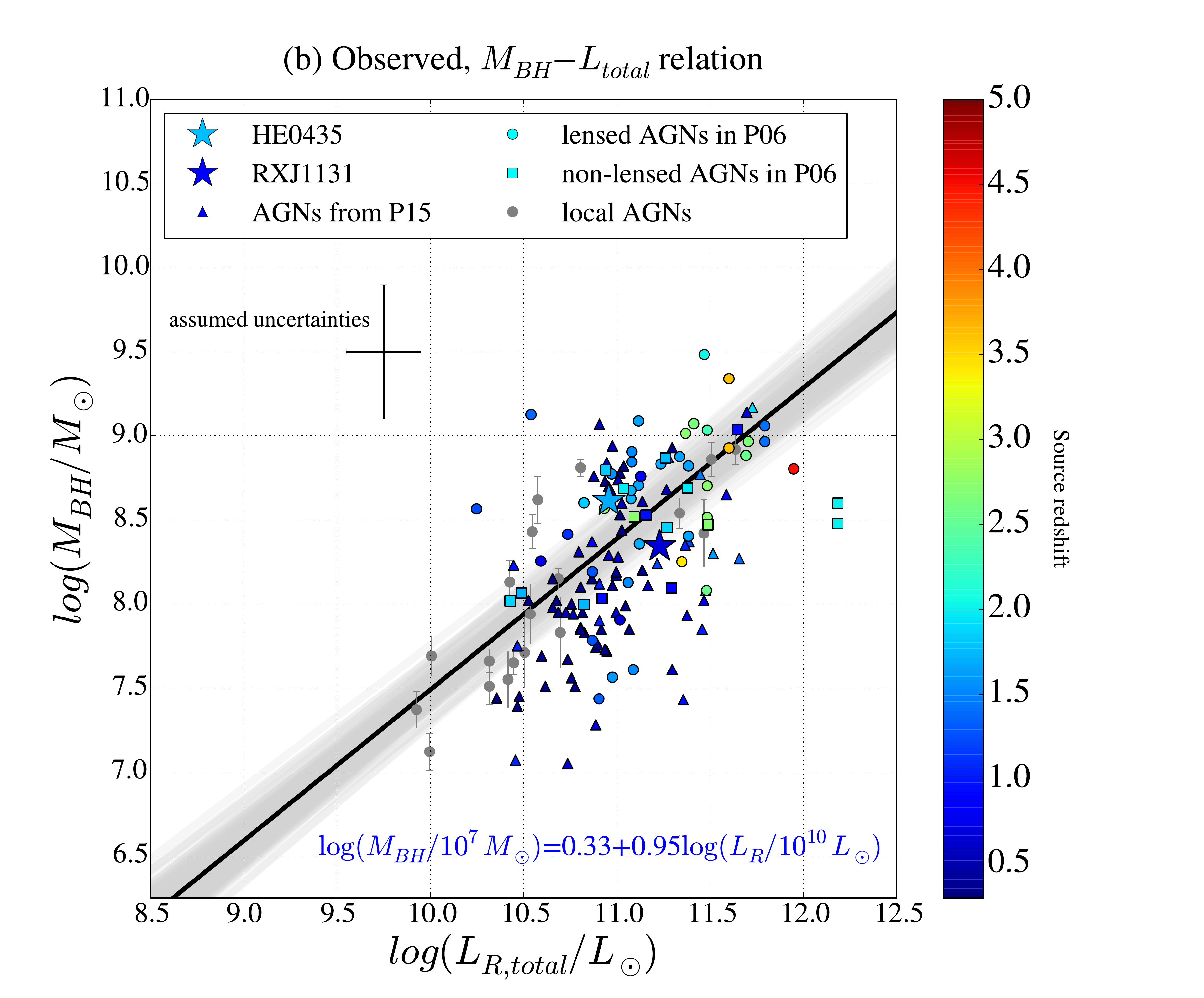}}\\
 \hspace{-5.5em}
{\includegraphics[trim = 0mm 0mm 90mm 0mm, clip, height=0.6\textwidth]{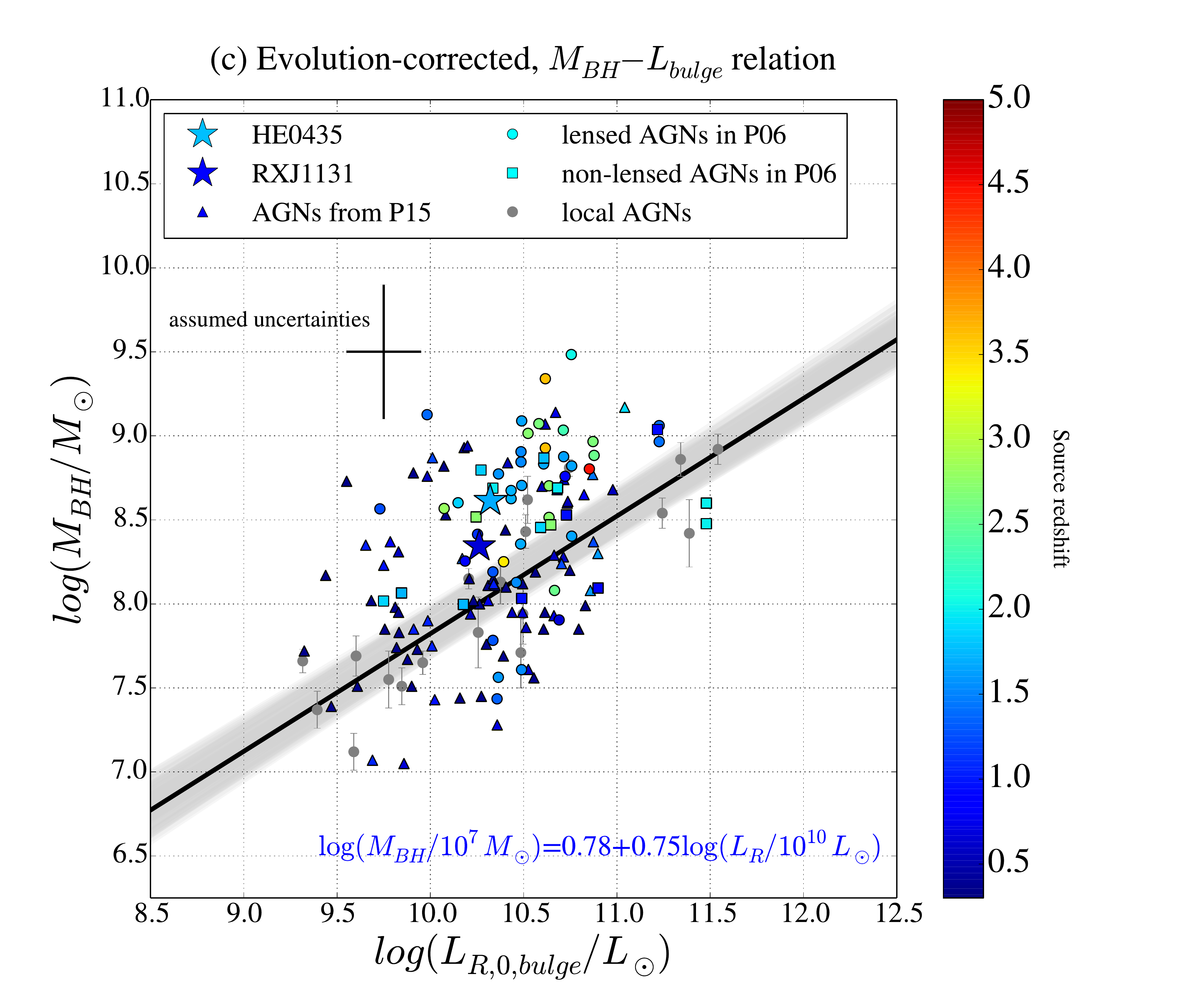}}& 
{\includegraphics[trim = 46mm 0mm 0mm 0mm, clip, height=0.6\textwidth]{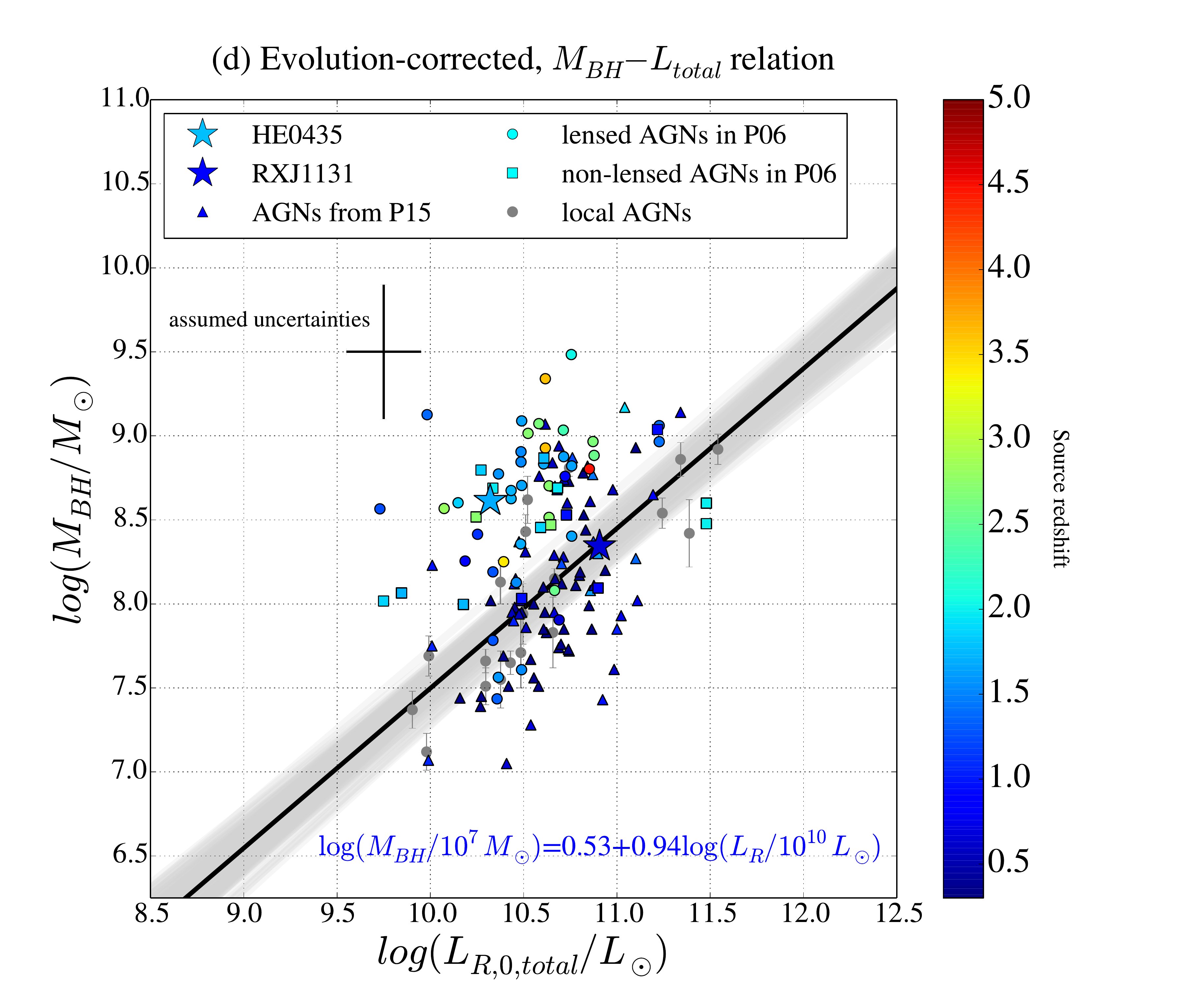}}\\
\end{tabular}
\caption{\label{fig:ML} Illustration of observed (top) and
  evolution-corrected (bottom) correlations of \mbh-\lbulge (left) and
  \mbh-\ltot(right). For distant AGNs, the redshifts are
  color-coded. The local data and their linear fitting (using an MCMC
  process) are colored in gray (1-$\sigma$ region) with the
  best-fitted coefficients in blue color.  We use the star symbol to
  highlight our new lensed-based measurements of HE0435 and
  RXJ1131. The total uncertainty for \lhost\ and \mbh\ of distant AGNs
  are adopted to be 0.2 dex ($\sim$ 0.5 mag) and 0.4 dex,
  respectively.}
\end{figure*}

%
The \mbh-\lbulge\ and \mbh-\ltot\ relation defined by our samples are
shown in Fig.~\ref{fig:ML}, panel (a) and (b). There is a clearly
positive correlation between \mbh\ and \lhost\ as in local
samples. For a direct comparison to local samples, we fit the local
\mbh-\lhost\ as:
\begin{eqnarray}
\label{eq:ml_relation}
\log \big( \frac{\mathcal M_{\rm BH}}{10^{7}M_{\odot}})= \alpha + \beta \log(\frac{L_R}{10^{10}L_{\odot}}).
\end{eqnarray} 
Using a Markov Chain Monte Carlo (MCMC) process we derive
$\alpha=0.68 \pm 0.18$; $\beta=0.74 \pm 0.09$ for the \mbh-\lbulge\
and $\alpha=0.33 \pm 0.22$; $\beta=0.95 \pm 0.15$ for the \mbh-\ltot,
with intrinsic scatter $\sigma_{\rm int}\sim 0.25$ for both of them.
Consistent with previous work (e.g. P06, P15), the observed
correlation at high redshift is nearly identical to the local. This is
perhaps surprising, considering that both the black hole mass and host
galaxy luminosity are expected to evolve over cosmic time. For
example, in a minimal evolution toy model, the elliptical galaxies and
their black hole are formed at high redshift and evolve passively
thereafter. Thus, we expect \lhost\ to fade over time, owing to aging
stellar populations. To allow a direct comparison to the local
samples, we considered this scenario in the next section.

\subsection{The passive evolution-corrected \mbh-\lhost\ relation}
\label{pas_ml}
In order to test the passive evolution scenario, we correct the
observed \lhost\ to its expected value at $z=0$ by accounting for the
aging of the stellar populations. It has been shown that the evolution
of the mass-to-light ratio of early-type galaxies can be {\it
  effectively} described as that of a single burst stellar population
formed at appropriate redshifts \citep[e.g.,][]{Tre++05}.  In order to
represent the uncertainty in the star formation history we consider a
range of single burst models formed at $z_f$ equals to 2, 3 and 5\footnote{
Stellar evolution is calculated with {\sc Galaxev}
  \citep{bruz2003}, based on Padova-1994 stellar evolutionary tracks,
  assuming Salpeter IMF, Solar metallicity, and no dust attenuation.}.
We choose to parametrize the evolution with the functional form
$d{\rm mag}_{\rm R}=\delta_m d\log(1+z)$, i.e.
\begin{eqnarray}
d\log L_{\rm R}/d\log(1+z)=\delta,
\end{eqnarray} 
with $\delta=-\delta_m/2.5$, so that
\begin{eqnarray}
\label{eq:L_relation}
\log(L_{R,0})=\log(L_{R}) - \delta \log (1+z).
\end{eqnarray} 

For this parametrization, we derive that $\delta_m \simeq -3.7\pm0.2$
(i.e. $\delta = 1.48\pm0.08$) provides a good representation of typical star formation histories.

\begin{figure}
\centering
 \hspace{-2.5em}
\includegraphics[height=0.44\textwidth]{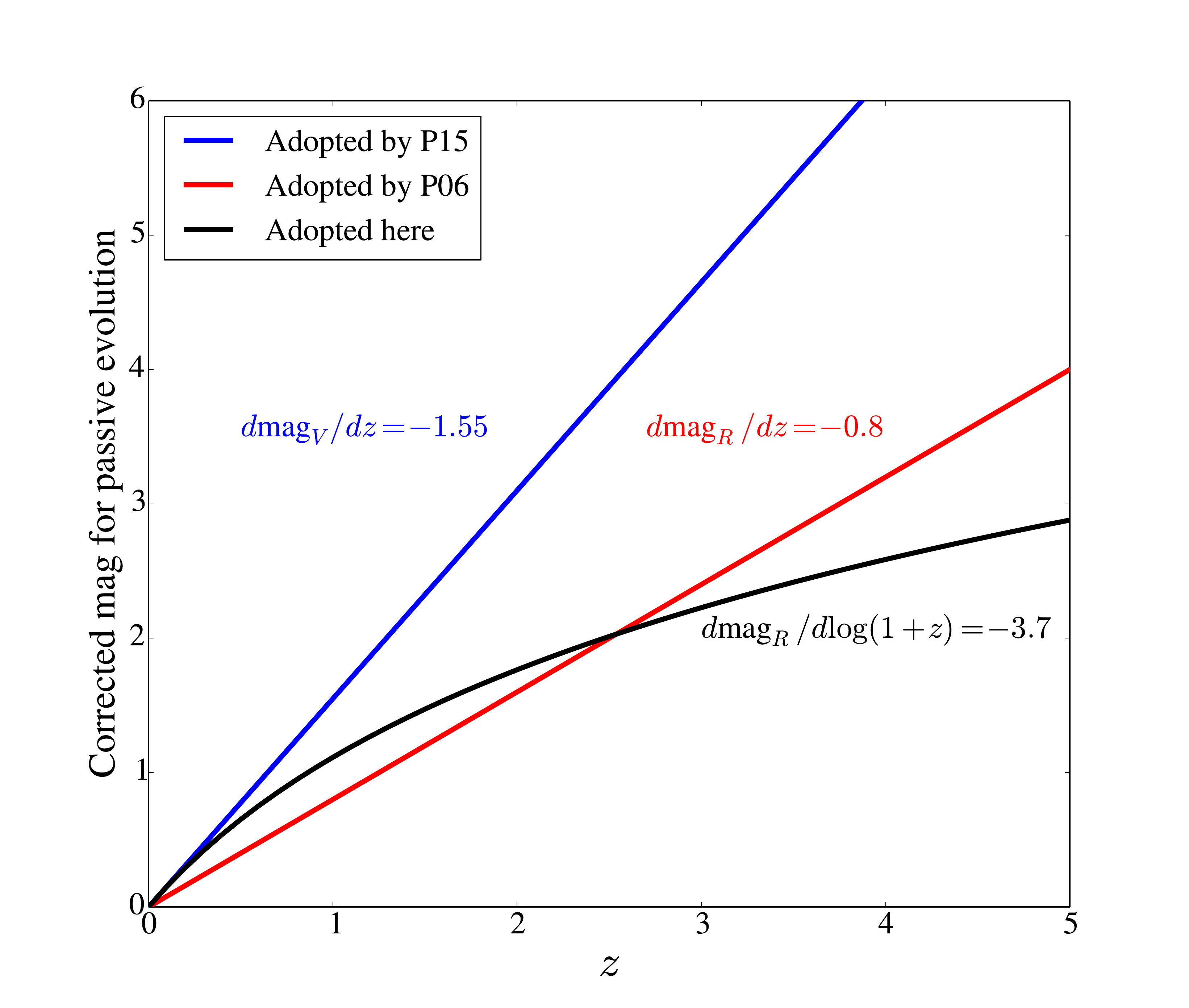}
\caption{\label{fig:dmag_com} Illustration of the comparison of the
  passive evolution correction adopted by P15, P06 and in this work.
  Note that all the samples in P15 are at low redshift
  ($z \lesssim 1$).  Thus, the $d{\rm mag}_V/dz \simeq -1.55$ is derived by
  assuming $z_f=2$ which is appropriate at these redshifts.  P06
  adopted $d{\rm mag}_R/dz \simeq -0.8 $ by assuming $z_f=5$.}
\end{figure}

This formalism is more accurate when considering a broad range in
redshift with respect to adopting a single slope as a function of
$d{\rm mag}/dz$ as done by P06 and P15.  For a direct comparison, we also
plot the passive evolution correction as a function of redshift in
Fig.~\ref{fig:dmag_com}.  Note that our chosen functional form
describes well the P15 form at $z<1$ and the P06 form at $z\sim3$
redshift, while avoiding the extreme corrections at very high-$z$
implied by previous parametrizations. Furthermore, our chosen
functional form facilitate the analysis of the \mbh-\lhost\ evolution
in the following way. Combining Eq.~\ref{eq:ml_relation} with the
passive evolving correction, i.e. Eq.~\ref{eq:L_relation},
and adding $\gamma'$ term which describes the evolution of the
correlation between \mbh\ and observed \lhost, leads to the following
formalism:
\begin{eqnarray}
\label{eq:cor_relation}
\log \big( \frac{\mathcal M_{\rm BH}}{10^{7}M_{\odot}})&~=~& \alpha + \beta \log(\frac{L_{R,0}}{10^{10}L_{\odot}}) \nonumber\\
&~+~& \beta \delta \log (1+z)
+\gamma' \log (1+z).
\end{eqnarray} 
In this equation, $\beta \delta$ represents the effects of passive evolution.
The evolution at fixed present-day luminosity is given by
$\gamma=\gamma'+\beta\delta$.
In this way the effects of the passive evolution correction can be
easily separated and a different passive evolution model can be
applied to the data, if desired. In our specific case, since we
derived $\beta = 0.74 \pm 0.09$ for the \mbh-\lbulge\ relation, the
passive evolution term corresponds to approximately
$\beta \delta = 0.74 \times 1.48 \sim 1.0$, neglecting the effects of
scatter and errors.  Likewise, the passive evolution term is
$\beta \delta = 0.95 \times 1.48 \sim 1.4$ for the \mbh-\ltot\
relation.  

The resulting \mbh-\lhost\ relation after applying the passive
evolution correction is shown in Fig.~\ref{fig:ML}, panels (c) and
(d).  Clearly, after the correction, the high redshift samples are
offset with respect to the local samples, indicating a tendency of BH
in the more distant Universe to reside in less luminous hosts at fixed
\mbh. This tendency is consistent with previous work, and also
consistent with the studies of the \mbh-$\sigma_*$ (stellar velocity
dispersion) and \mbh-\mstar\ (stellar mass) correlations, which do not
require correction for passive evolution
\citep{TMB04,Woo++06,Woo++08,Ben++11}.

\begin{figure*}
\centering
\begin{tabular}{c c}
 \hspace{-3.4em}
\subfloat[Offset for \mbh-\lbulge\ relation using entire sample]
{\includegraphics[width=0.59\textwidth]{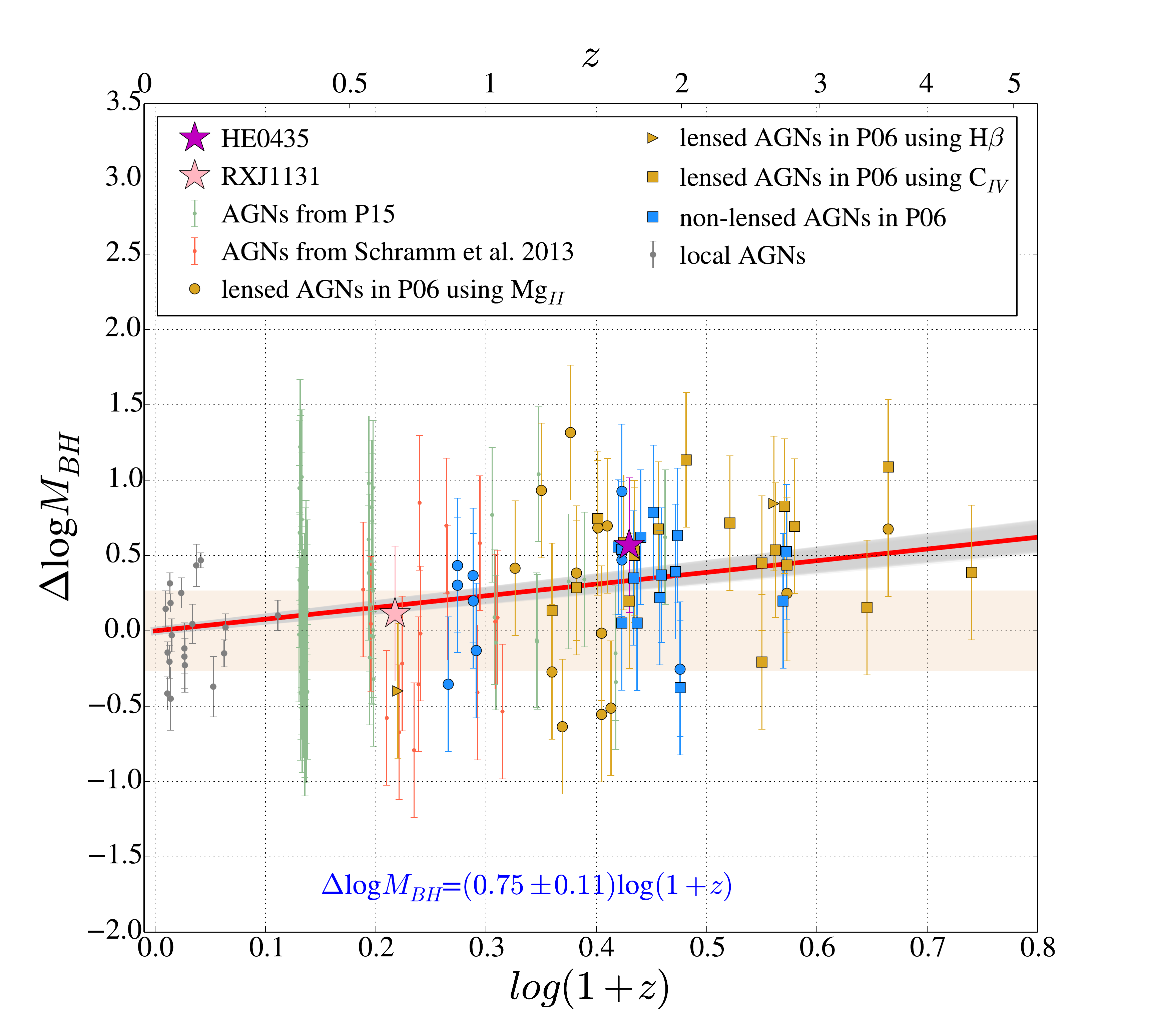}}&  \hspace{-5.2em}
\subfloat[Offset for \mbh-\ltot\ relation using entire sample]
{\includegraphics[width=0.59\textwidth]{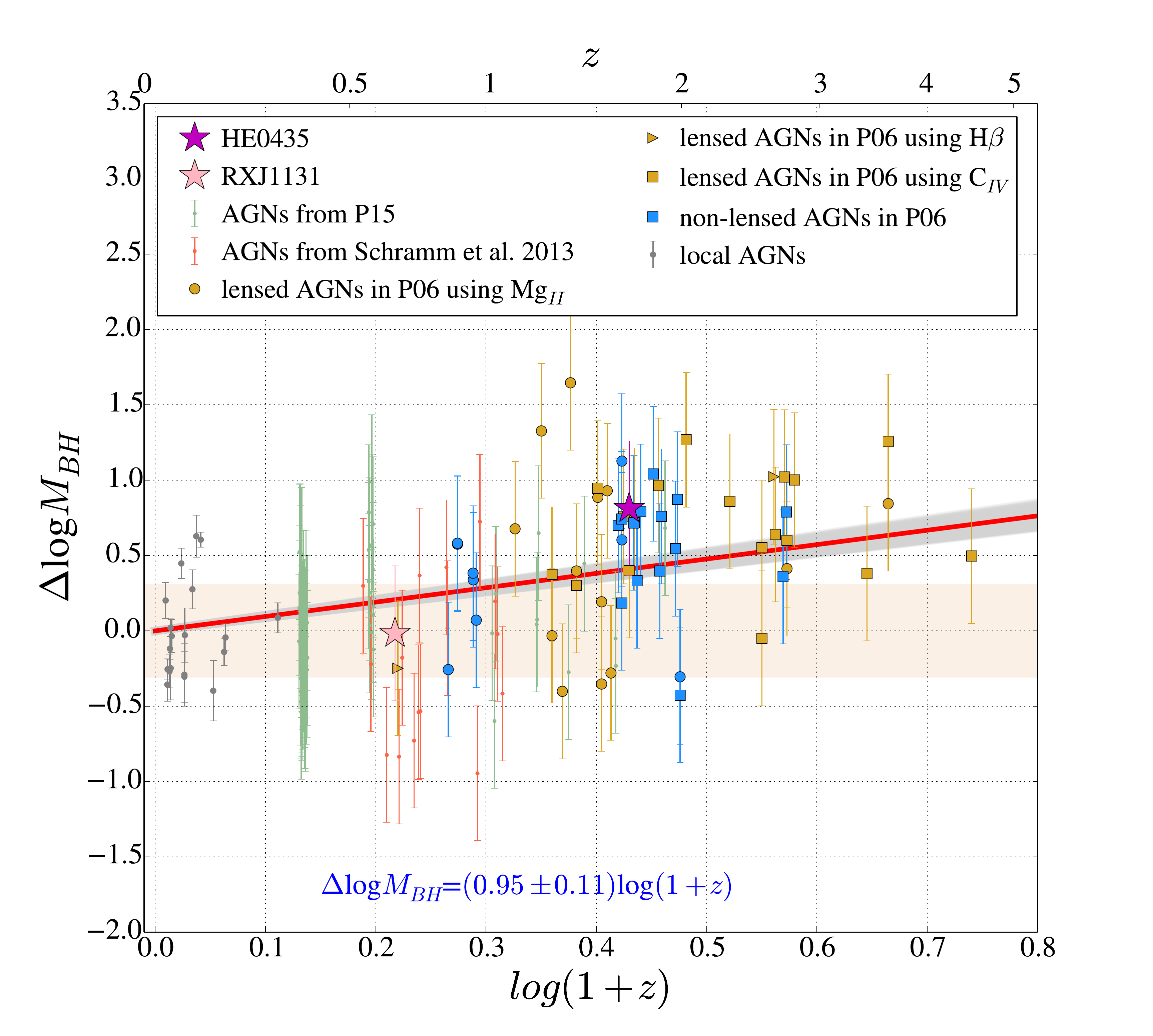}}\\
 \hspace{-3.4em}
 \subfloat[Offset for \mbh-\lbulge\ relation using reduced sample]
{\includegraphics[width=0.59\textwidth]{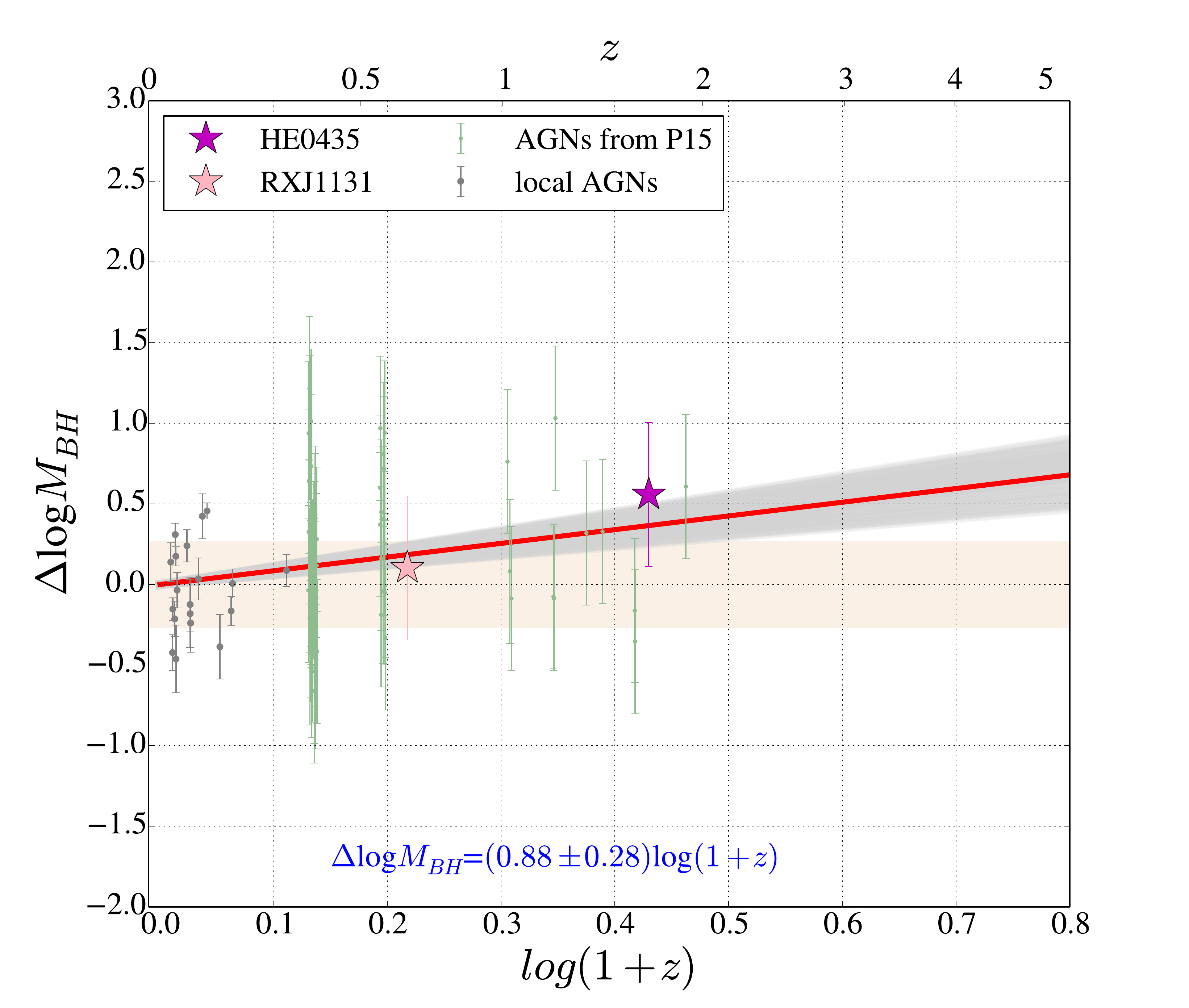}}&  \hspace{-5.2em}
\subfloat[Offset for \mbh-\ltot\ relation using reduced sample]
{\includegraphics[width=0.59\textwidth]{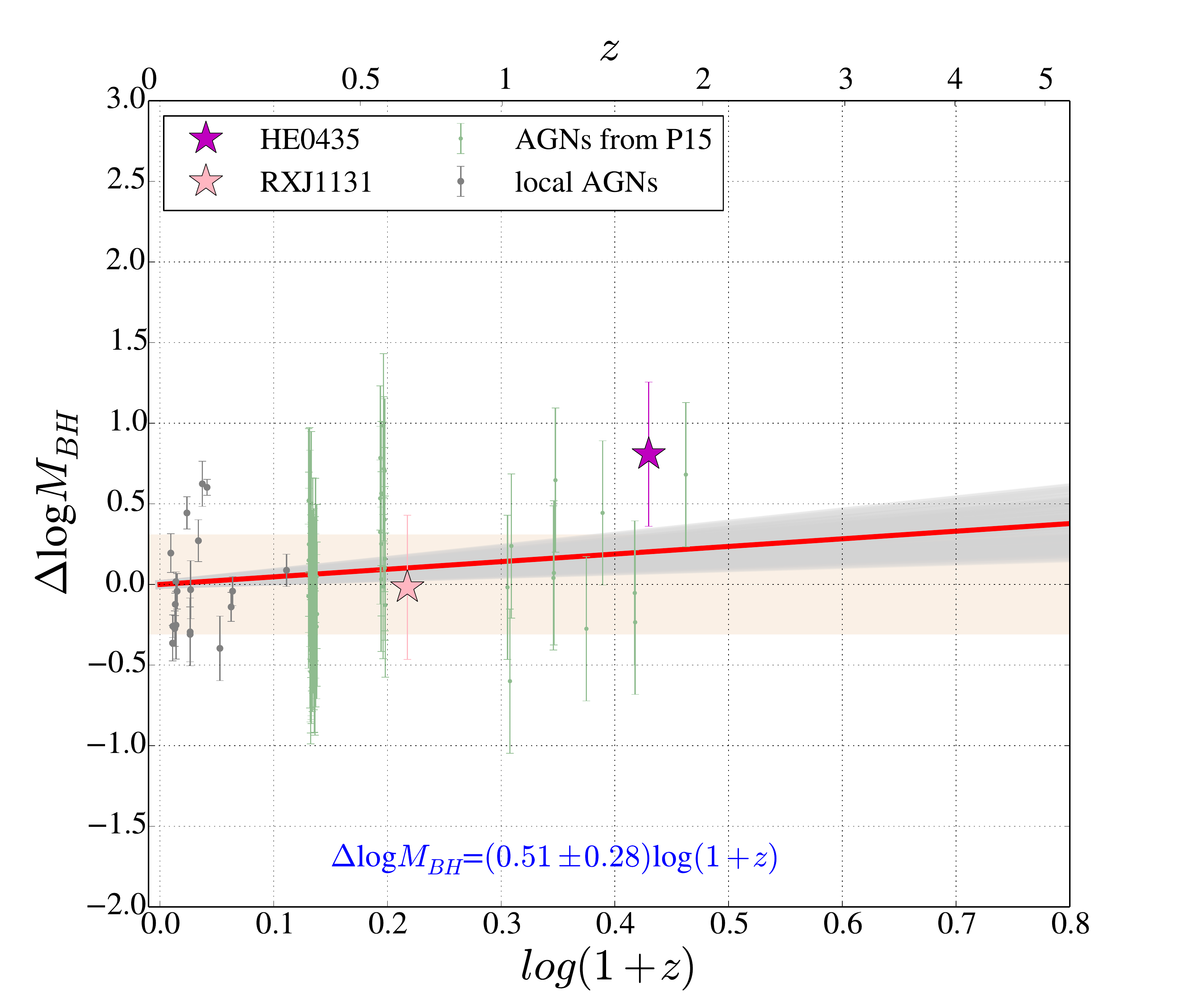}}\\
\end{tabular}
\caption{\label{fig:MLz} Illustration of the offset in log\mbh\ for a
  given \lbulge\ (left) and \ltot (right) as a function of redshift,
  after passive evolution correction.  Top panels corresponds to the
  fitting using the whole sample. We also highlight the subsamples
  from SS13.  Bottom panels corresponds to the fitting excluding the
  samples from P06 and SS13.  The red solid line represents the
  best-fit trend for all distant objects as a functional of
  $\Delta \log$\mbh $ = \gamma \log (1 + z)$, with the 1-$\sigma$
  region confidence range shaded in grey.  The orange band is the
  intrinsic scatter of local linear relation.  }
\end{figure*}

We fit the offset in black hole mass at fixed passively evolved
luminosity as a function of redshift in the form:

\begin{eqnarray}
\label{eq:offset}
\Delta \log \mathcal M_{\rm BH}= \gamma \log (1 + z),
\end{eqnarray} 
where
$\Delta \log \mathcal M_{\rm BH} = \log \big( \frac{\mathcal M_{\rm
    BH}}{10^{7}M_{\odot}})
-\alpha-\beta\log(\frac{L_{R,0}}{10^{10}L_{\odot}}), $ and obtain
$\gamma = 0.75 \pm 0.11$ for the \mbh-\lbulge\ and
$\gamma= 0.95 \pm 0.11$ for the \mbh-\ltot, as shown in
Fig.~\ref{fig:MLz}, panels (a), (b).  We also obtain
$\gamma' = -0.14 \pm 0.11$ for the \mbh-\lbulge\ and
$\gamma'= -0.26 \pm 0.12$ for the \mbh-\ltot , when not taking into
account the passive evolving correction.  As expected, the difference
$\gamma-\gamma'$ is consistent with the effects of the passive
evolving correction, i.e. $\beta \delta \sim 1.0$ for the
\mbh-\lbulge\ and $\beta \delta \sim 1.4$ for the \mbh-\ltot.

We conclude by noting that this fit does not take into account
selection effects, which are discussed in the next section.

\subsection{Selection effects}

From Fig.~\ref{fig:ML}, we can see that at high redshift we
preferentially study systems with the larger \mbh\ and \ltot. This is
expected as observational samples tend to be flux limited and thus
favor the high luminosity tail (and hence typically high \mbh) of the
distribution. Like many other instances in astronomy, it is essential
to take into account the selection function when estimating the
evolution of the black hole mass host galaxy correlations
\citep{Tre++07,Lau++07,Ben++11,S+W14,Park15}.

Following P15, we take selection effects into account by using a Monte
Carlo simulation method based upon the methodology introduced by
\citet{Tre++07} and \citet{Ben++10}. The simulated samples are
generated from a combination of the local active BH mass function from
\citet{Schu++2010} and the local \mbh-\lhost\ relation from
\citet{Ben++10} with Gaussian random noise added as a function of the
two free parameters $\gamma$ and intrinsic scatter of the correlation
$\sigma_{\rm int}$.  Note that the scatter is assumed to be
independent of redshift in our description. For each object, the
likelihood of the observed \mbh\ with a given \lhost\ is calculated
from the simulated sample at the given $\gamma$ and
$\sigma_{\rm int}$, and taking into account whether the object would
be selected or not based on our sensitivity.  Finally, by adopting
uninformative uniform (flat) prior or lognormal prior from
\citet[][$\sigma_{\rm int}=0.21\pm0.08$]{Ben++10}, the posterior
distribution function of $\gamma$ and $\sigma_{\rm int}$ is evaluated.
Selection effects are modelled in the same way for the
  lensed-quasar sample, neglecting any second order effects related to
  lensing magnification.  We note however, that these effects are
  small \citep{C+C16} and magification-related biases should affect
  the quasar and host galaxy in a similar manner, thus moving objects
  mostly along the \mbh-\lhost\ correlation and not away from it.

\begin{figure*}
\centering
\begin{tabular}{c c}
 \hspace{-4em}
 \subfloat[\mbh-\lbulge, flat prior]
{\includegraphics[width=0.55\textwidth]{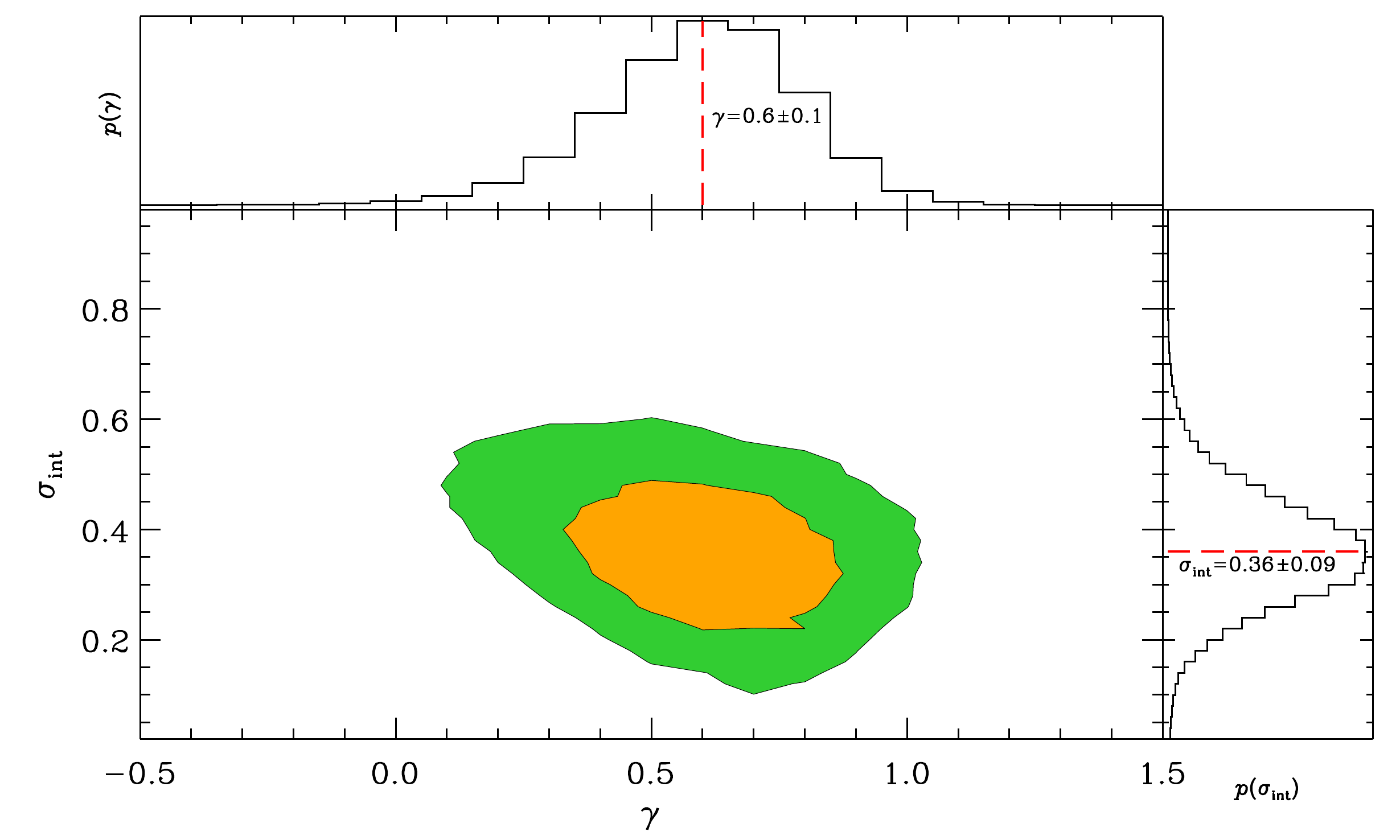}}&  \hspace{-2em}
 \subfloat[\mbh-\lbulge, lognormal prior]
{\includegraphics[width=0.55\textwidth]{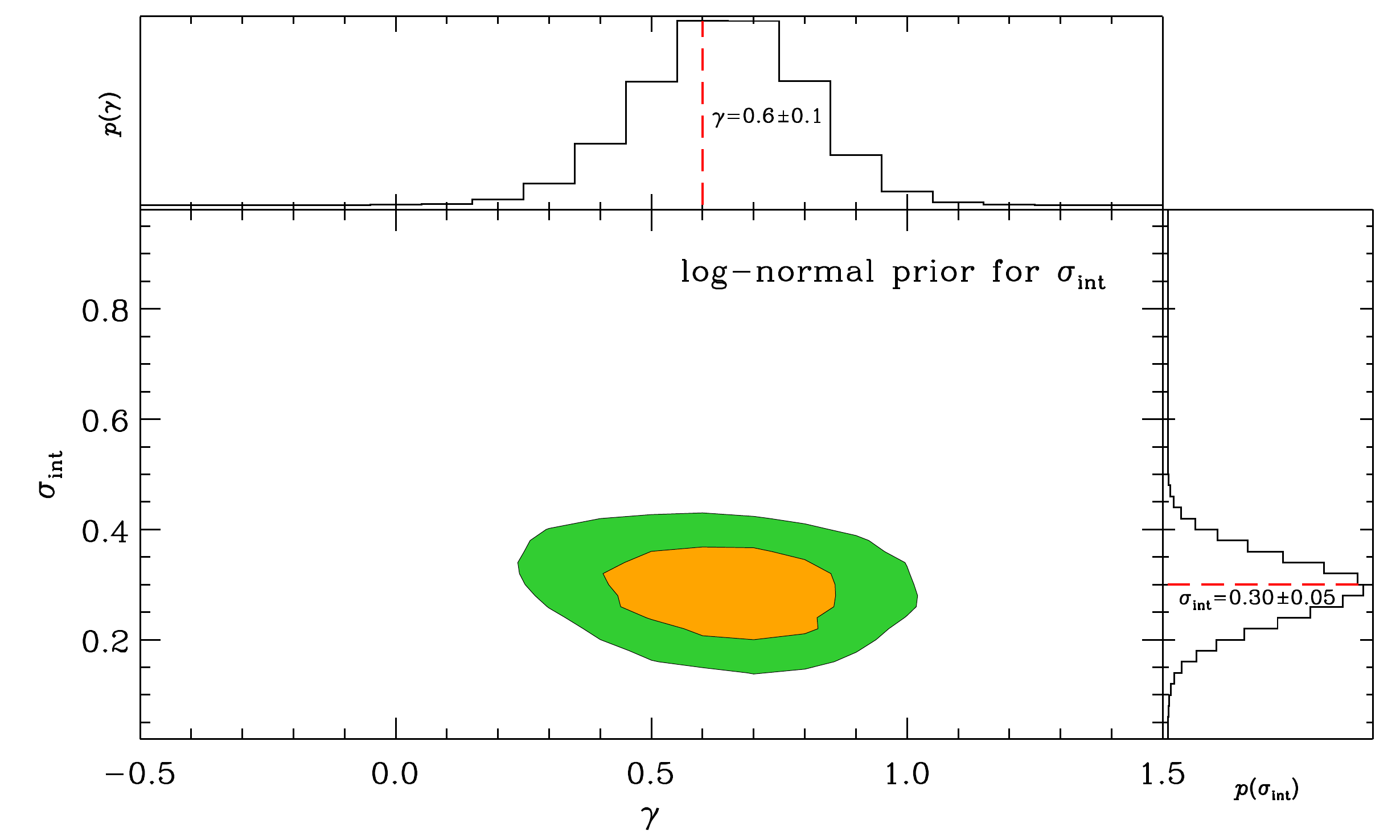}}\\
 \hspace{-4em}
 \subfloat[\mbh-\ltot, flat prior]
{\includegraphics[width=0.55\textwidth]{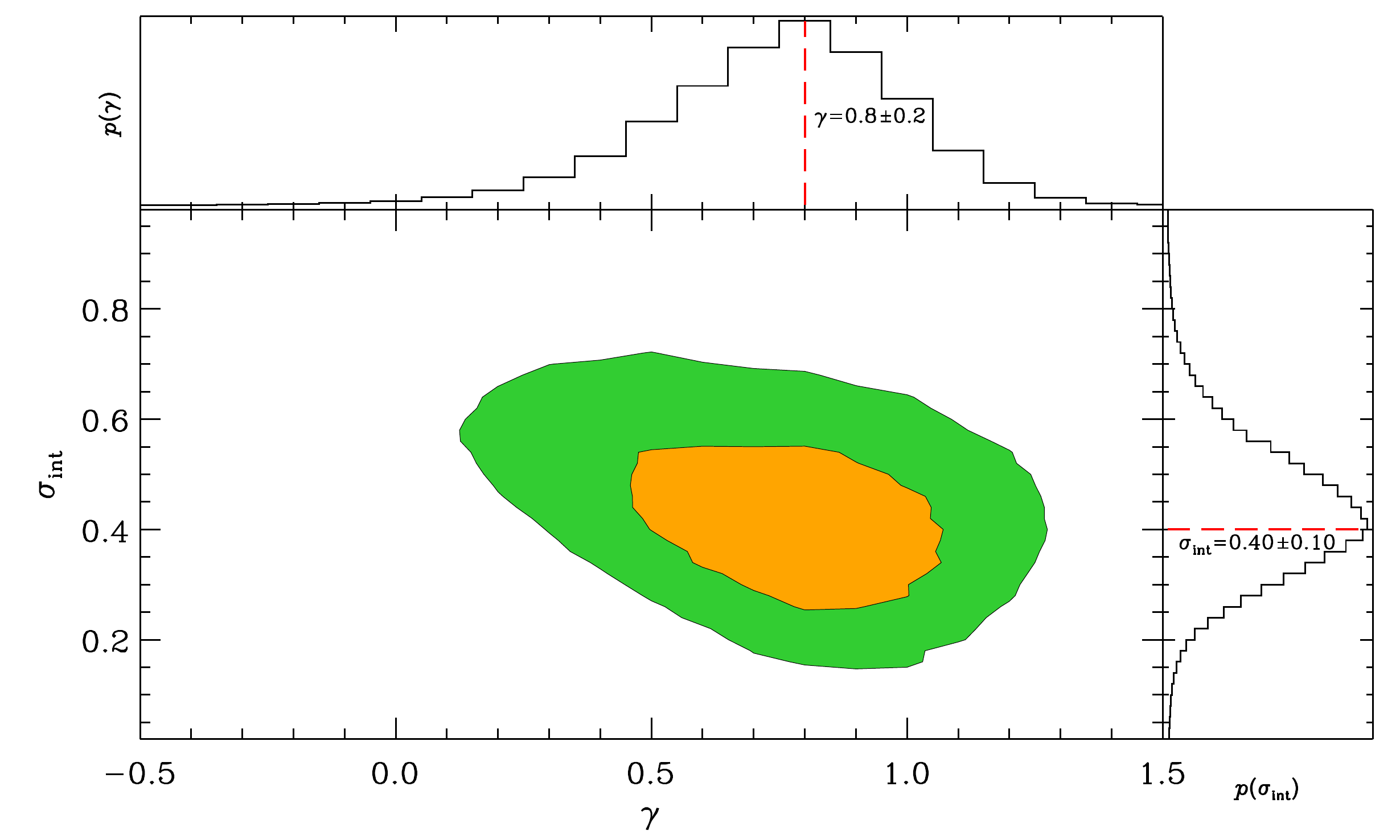}}&  \hspace{-2em}
 \subfloat[\mbh-\ltot, lognormal prior]
{\includegraphics[width=0.55\textwidth]{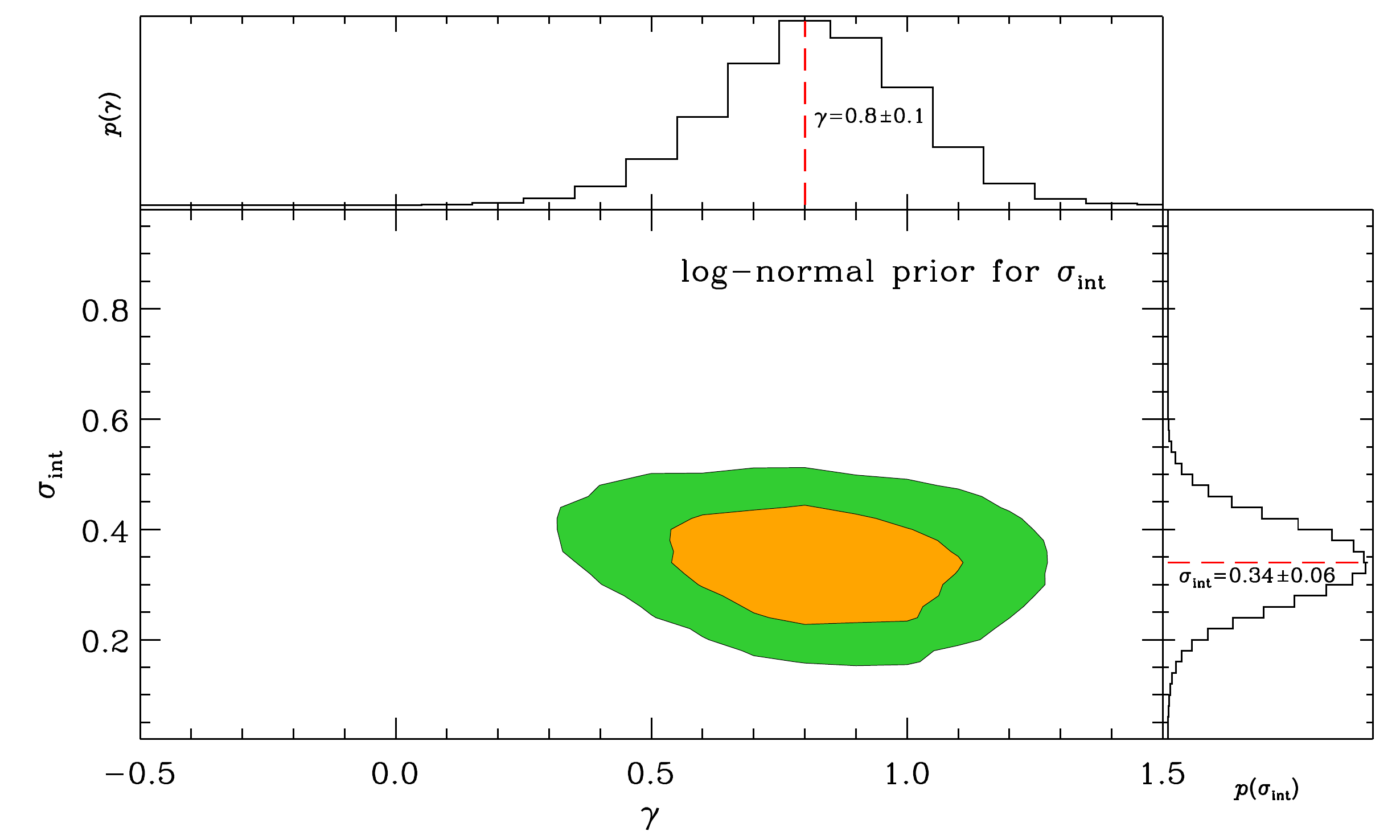}}\\
\end{tabular}
\caption{\label{fig:slt_all_eff} Posterior distribution function given
  the entire dataset for a model with evolution in the form
  $\Delta \log$\mbh $ = \gamma \log (1 + z)$ with intrinsic scatter
  $\sigma_{\rm int}$, taking into account selection effects.  The
  \mbh-\lbulge (top) and \mbh-\ltot (bottom) correlations with flat
  (left) and lognormal prior (right) are shown.  }
\end{figure*}

Taking into account selection effects, the results of the inference
are shown in Fig.~\ref{fig:slt_all_eff}.  The fitted values of
$\gamma$ are $0.6 \pm 0.1$ (\mbh-\lbulge) and $0.8 \pm 0.2$
(\mbh-\ltot), almost independent of the choice of prior.  These values
are consistent with the previous inference in Section \ref{obs_ml} and
\ref{pas_ml}.

Interestingly, the intrinsic scatter of the correlations is found to
be consistent with typical values inferred for local samples ($0.3-0.4$
dex). This result is consistent with the hypothesis that well defined
correlations exist at the redshifts probed by our sample, and indicates
that we have not significantly underestimated our errors at high-$z$.
It would be beneficial to study how the selection bias changes as a
function of some key factors such as the values of \lhost\ and \mbh, the level
of the uncertainties and the redshift distribution of the samples. However,
this topic is trivial in this study as we obtained consistent inference by
either or not talking selection effects into account.
Moreover, to study this relation quantitively requires considerate tests and simulations.
Thus, we leave it in the future study.

\section{Discussion}
\label{sec:disc}

In this section, we first estimate the importance of potential
systematic errors in \S~\ref{disc:syserror}. Then, we carry out a
detailed comparison with previous observational work in
\S~\ref{disc:compare}.  Finally we discuss how our measurements fit
into our understanding of galaxies and BHs co-evolution in \S~\ref{diss_imp}.

\subsection{Systematic errors}
\label{disc:syserror}
We have combined our new measurements with ones taken
from the literature in order to increase the sample size and reduce
statistical uncertainties. Even though we have restricted our analysis
to the samples that have been analyzed in the most similar manner to
our new data and we have cross-calibrated the black hole mass
estimators, there are still some residual differences.

First, P06 
obtained the luminosity of one galaxy by combined the fluxes together,
even though some
of them may include a disk component (e.g. RXJ1131).
According to
morphological studies of AGN host galaxies, the fraction of
spiral/elliptical hosts of AGN is approximately one third
\citep{Kocevski++12}, with the exact value depending on \mbh\ and
luminosity.  Thus it is possible that P06 overestimates the bulge
component of some of the host galaxies. The total luminosity should be
less affected by this bias, even though not completely immune.

Furthermore, the subsample by SS13 included in the compilation by P15,
was X-ray selected as opposed to optically selected like the rest of
the non-lens sample (some of the lenses are radio-selected). This
difference in selection could potentially lead to a systematic
difference between the two samples.

\begin{figure*}
\centering
\begin{tabular}{c c}
\hspace{-4em}
 \subfloat[\mbh-\lbulge, flat prior]
{\includegraphics[width=0.55\textwidth]{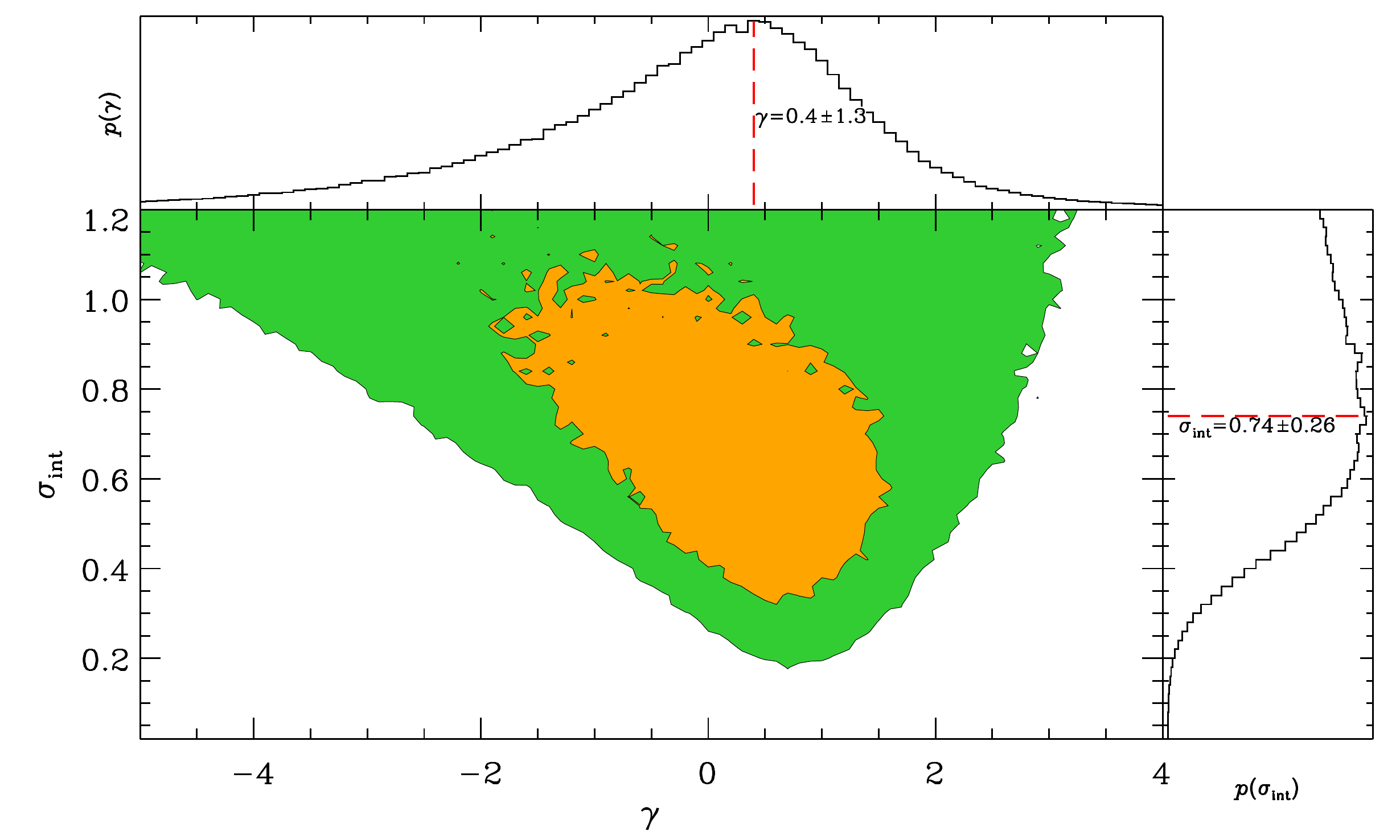}}&  \hspace{-2em}
 \subfloat[\mbh-\lbulge, lognormal prior]
{\includegraphics[width=0.55\textwidth]{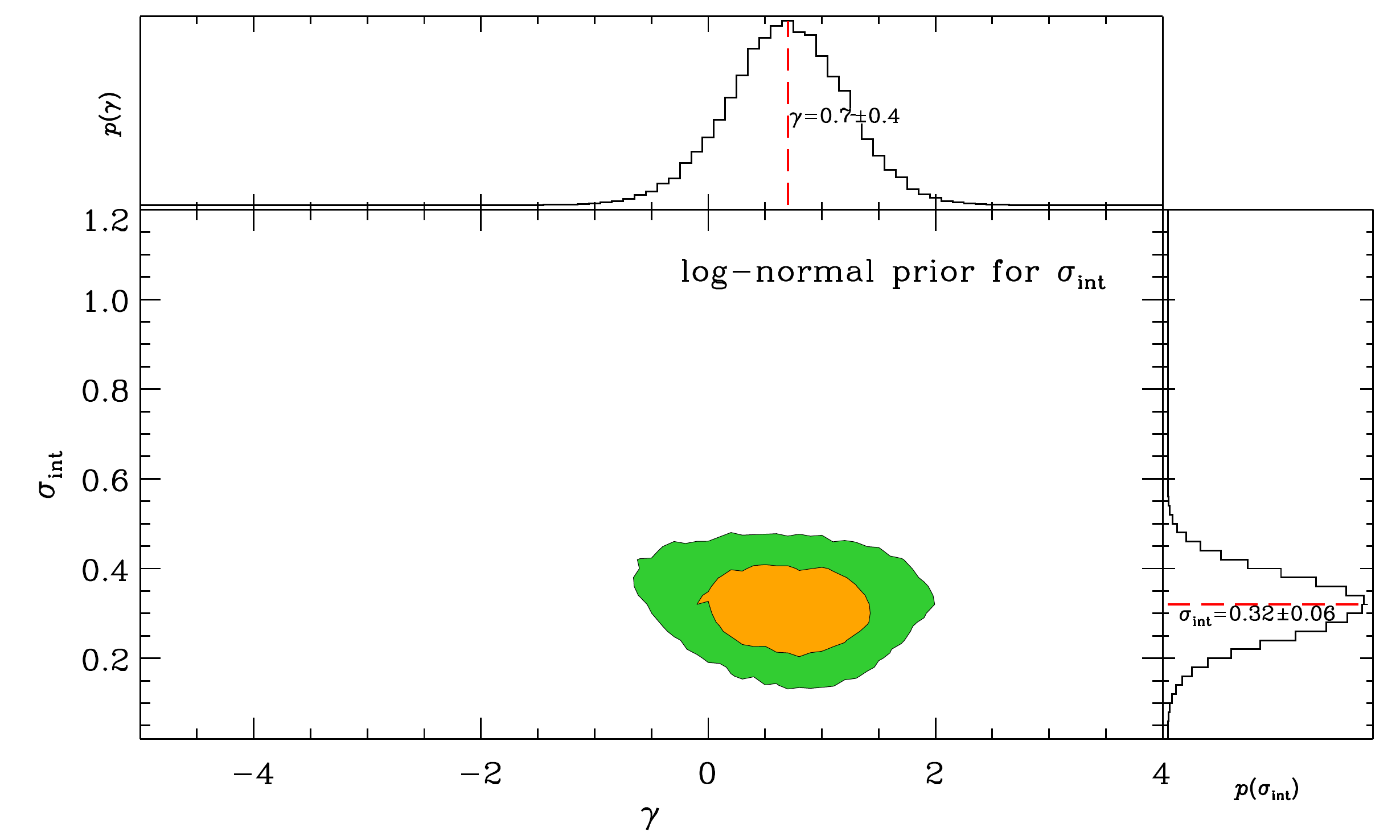}}\\
\hspace{-4em}
 \subfloat[\mbh-\ltot, flat prior]
{\includegraphics[width=0.55\textwidth]{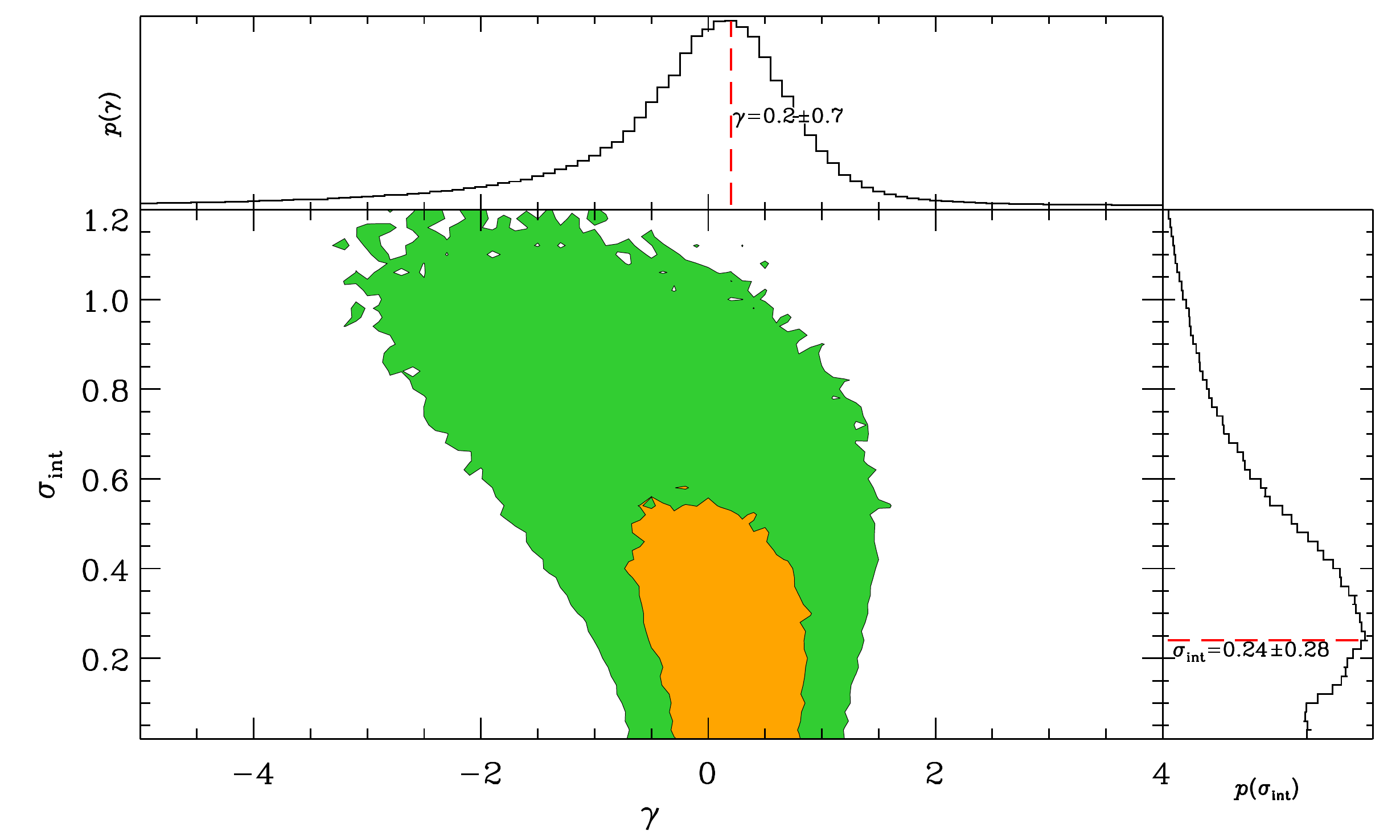}}&  \hspace{-2em}
 \subfloat[\mbh-\ltot, lognormal prior]
{\includegraphics[width=0.55\textwidth]{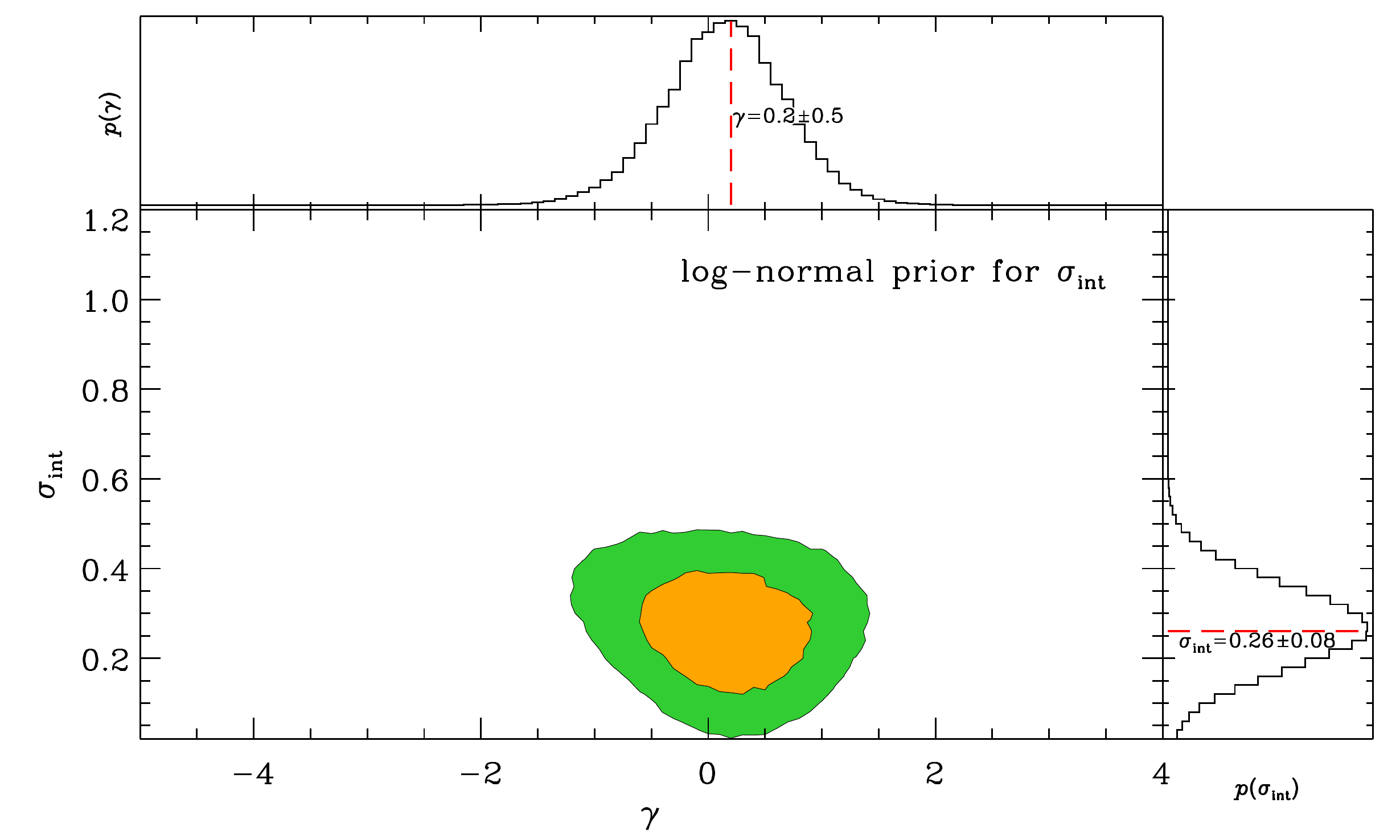}}\\
\end{tabular}
\caption{\label{fig:slt_red_eff}
Same as Fig.~\ref{fig:slt_all_eff}, using the {\it reduced} sample.
}
\end{figure*}

In order to estimate these systematic uncertainties, we repeat the
analysis by excluding the P06 and SS13 samples. This reduced sample
will have significantly less statistical power, owing to the reduced
size and redshift coverage, but should be more robust with respect to
the systematic uncertainties discussed above. Given this reduced
sample, we obtain $\gamma = 0.88 \pm 0.28$ for the \mbh-\lbulge\ and
$\gamma= 0.51 \pm 0.28$ for the \mbh-\ltot, as shown in
Fig.~\ref{fig:MLz}, panel (c), (d).  Moreover, we use the same
approach to study the selection effects and obtain the consistent
inference, as illustrated in Fig.~\ref{fig:slt_red_eff}.  Even though
as expected the uncertainties are larger than for the full sample, the
results are statistically mutually consistent at 1-$\sigma$ level.  To
facilitate the comparison between different $\gamma$, we summarize our
inference in Tab.~\ref{comp_gama}.  We conclude that our inferred
trends are not dominated by systematic differences between the
samples, and systematic uncertainties of this kind are smaller than
the random ones.

In this work, \mbh\ estimates are derived using the \Civ, \Mgii\ and \Hb\
emission lines. However, the \Civ\ and \Mgii\ lines are usually in
outflow \citep{Baskin+2005, Richards:2011p8131, Denney2012} and
therefore may not be dominated by the gravity of the central \mbh\ and
result in biased \mbh\ estimates, especially for the \Civ\ line
\citep{Trakhtenbrot2012}.  Following \citet{McG++08} the potential
bias has been mitigated by cross-calibrating the \mbh\ estimates based
on the different lines. As a further sanity check, we fitted the
$\gamma$ using only \Hb -based samples. We note that this \Hb\ sample
is very smilar to the subsample excluding P06 and SS13, and in fact
the results are similar ($\gamma = 1.10 \pm 0.36$ for the \mbh-\lbulge\
and $\gamma= 0.7 \pm 0.37$ for the \mbh-\ltot).  We conclude that any
potential residual bias related to the use of lines other than \Hb\ is
smaller than statistical uncertainties or biases related to sample
selection.

\begin{table}
\setcounter{table}{3}
\centering
    \caption{The summary for the different inference of $\gamma$.}\label{comp_gama}
    \resizebox{9cm}{!}{
     \begin{tabular}{lccc}
     \hline
     Sample & Account for& \mbh-\lbulge&  \mbh-\ltot \\
     &selection effects? &&\\
     \hline\hline
Entire & No & $ 0.75 \pm 0.11$ & $ 0.95 \pm 0.11$ \\
Entire & Yes$^{a}$ & $0.6\pm0.1$ & 0.8$\pm0.1$ \\
Exclude P06, SS13& No & $0.88\pm0.28$ & 0.51$\pm0.28$\\
Exclude P06, SS13& Yes & $0.7\pm0.4$ & 0.2$\pm0.5$\\
     \hline
     \end{tabular}}
     \begin{tablenotes}
      \small
      \item
$^a$: Using lognormal prior.\\
     \end{tablenotes}
\end{table}

\subsection{Comparison with previous work}
\label{disc:compare}

P15, using a sample of 79 active galaxies, inferred the following
evolutionary trends: $\gamma = 0.9 \pm 0.7$ for the \mbh-\lbulge\ and
$\gamma = 0.4 \pm 0.5$ for the \mbh-\ltot. These are consistent with
our inference, although their uncertainties are much larger, owing to
the smaller sample size and reduced high redshift coverage. A similar
result was obtained by P06, where they found that the ratio between
\mbh\ and \mstar\ was $\sim 4$ times larger at $z\sim 2-4$ than today
(i.e.  $\gamma \sim 0.8-1.2$). The consistency between their
measurements and ours are expected since the overall samples in this
work are mostly composed of the samples by P15 and P06, even though
there are some differences in the rest-frame bands chosen for
photometry (we and P06 adopt rest-frame R, while P15 adopts rest-frame
V), in the passive evolution correction, and in the black hole mass
calibration.

The cosmic evolution of the \mbh-\lbulge\ relation is a topic of
intense debate in the literature.  Many works have reported an
evolutionary signal based on different relations including the
\mbh-\lhost\ \citep[e.g.][]{Tre++07, Ben++10}, the \mbh-\mstar\
\citep[e.g.][]{McL++06a, Jah++09, Dec++2010, Cisternas:2011p12133,
  Bennert11, Trakhtenbrot2015} and the \mbh-$\sigma_*$ \citep[e.g.][]{Woo++06, Woo++08}
correlations.  Nevertheless, other observational studies
\citep[e.g.][]{Shi++03, G+H05a, Komo+2007, Shen++2008} found no
evidence for evolution.
In \citet{Shankar2016}, they find serious biases in the \mbh-\mstar\ relation
and prove that $\sigma_*$ is more fundamental than any other variable.
However, in \citet{Shankar2009}, they show that there is no evolution in the \mbh-$\sigma_*$
relation once one accounts for the ages of local galaxies and the So\l tan argument.
Moreover, \citet{SW++2011, S+W14} concluded
that there is no statistically significant evidence for evolution once
these selection effects are taken into account and corrected.  Taking
a different approach, \citet{DeG++15} used the results of the
high-resolution numerical simulation {\sc MassiveBlackII} to compare
the observed and intrinsic evolution of the black hole mass host
galaxy correlations and reproduced the evolutionary trend of the
relation.  Consistent with other considerations, they also found that
the observed samples display steeper slopes than random ones,
suggesting the selecting effects can exhibit faster evolution than a
random sample.  Similarly, by generating Monte Carlo realizations of
the \mbh-$\sigma_*$ relation at $z = 6$, \citet{Volonteri++2011} also
found that due to selection bias the 'observable' subsample would
suggest an average positive evolution even when the intrinsic
correlations is characterized by no or negative evolution at high
redshift.  These studies highlight once again the importance of taking
selection effects into account.

Clearly, absence of evidence does not imply evidence of absence, and
one way to make progress is to improve the precision and accuracy of
the measurement. In our work, we attain much higher precision than
previous work owing to the enlarged sample, including lensed
quasars. Thanks to the large sample size, even when selection effects
are taken into account, the evolutionary trend is detected at high
significance ($\gamma\neq0$ at more than 5-$\sigma$).  However, using
a reduced sample by excluding the subsamples from P06 and SS13, we
obtain a smaller evolutionary trend, with larger uncertainties. These
results are consistent at 1-$\sigma$ level (see Tab.~\ref{comp_gama}),
and highlight the importance of studying larger sample of high
redshift lenses with the state-of-the-art data.

\subsection{Implication for the co-evolution of black holes and their host galaxies}
\label{diss_imp}

Our results are consistent with a scenario in which BHs in the distant
Universe typically reside in lower stellar mass galaxies than today,
assuming that the passively evolved luminosity tracks approximately
stellar mass \citep[see][for a consistent direct measurement based on
stellar mass determination]{Ben++11,SS13}.  In order to end up on the
local final relation, the stellar mass of the host galaxy would have
to grow faster than \mbh.

An interesting clue to the physical mechanism driving the evolution
could perhaps be found by comparing the inferred evolution for the
correlation between \mbh\ and the total host galaxy luminosity, and
that with the bulge luminosity. We found those two to be comparable
within the uncertainties. Previous work found the
\mbh-\mstar$_{\rm bulge}$ correlation to evolve somewhat faster than
\mbh-\mstar$_{\rm total}$, albeit at low statistical significance,
suggesting that one of the mechanisms at work is the build up of the
bulge component from stars in the disk \citep{Cro06}. It is difficult
to perform a direct comparison, because of the fact that the P06
sample did not attempt bulge-disk decomposition, and we have also
assumed a single passive evolution trend for the entire galaxy. Both
effects could potentially suppress the differences between the
evolution of the bulge and total luminosity with respect to the black
hole mass.

Also, our sample extends to much larger redshift than that of B11. One
possible explanation of this possible tension is that the dominant
evolutionary mechanisms changes with redshift.  At low redshift
($z\lesssim1$), the growth of the bulge is dominated by the secular
evolution with the redistribution of disk stars while at high redshift
($z\gtrsim1$), the growth is dominated by major mergers
\citep[see][for a similar conjecture]{Ben++10}.

To settle this issue, it is crucial to obtain high-quality data and
model large samples of lens systems, so that robust bulge to total
luminosity decompositions can be carried out. It would also be
beneficial to obtain multi-color data to estimate directly stellar
mass, and ideally stellar kinematic information to distinguish
pressure supported systems from rotationally supported ones.

Since the \lhost\ of ellipticals would not change when considering the
bulge and the total, we examine the offset using the sample limited to
spiral galaxies. In our sample, there are 9 local spirals and 41
distant spirals, excluding SS13. Fitting the offset with this
subsample, we obtain $\gamma=2.15\pm0.41$ and $1.18\pm0.41$ for
\mbh-\lbulge\ and \mbh-\ltot, respectively, which are larger than the
previous inference listed in Tab.~\ref{comp_gama} for the entire
sample.  This difference could suggest the spiral galaxies are
undergoing a more rapid evolution than the ellipticals in order to end
up on the local final relation. However, we caution that this result
should be taken with a grain salt, given the small sample size of the
local disk comparison sample.

\section{summary}
\label{sec:sum}

We presented a new measurement of the co-evolution of supermassive
black holes and their host galaxies.  First, we carried out a new
analysis of two strongly lensed quasars, \hequad\ and \rxjquad. By
using the state-of-the-art lens models by \citet{H0licow4} and \citet{Suy++13}, we
found that the host galaxies of HE0435 and RXJ1131 are well described
by an elliptical and spiral surface brightness density profile,
respectively. Then, we measured the host galaxy magnitude and tested
for potential biases by carrying out realistic simulations following
the procedure outlined by \citet{H0licow6}. We found that the bias of
our inference of \lhost\ is small (0.1$-$0.2 mags) and that we can
recover the host image precisely even if the host has multiple
components (see Fig.~\ref{fig:sim_1131}, panel (c)). We estimated \mbh\
by using a set of self-consistent single epoch estimators based on the
quasar emission line properties as measured by \citet{Sluse++2012}.

Second, we combined our measurements with the published ones
from the literature \citep{Pen++06qsob, Park15}, thus expanding our
sample to $146$ active galaxies up to $z=4.5$. We have taken care of
using self-consistent recipes to re-derive the black hole mass
estimates and convert all the luminosities self-consistently to the
rest-frame R-band.

Our main findings can be summarized as follows:

\begin{enumerate}
\item The observed correlations - without correction for evolution -
  are consistent with those observed in the local Universe.
\item The data are inconsistent with a passive evolution scenario. By
  correcting the host galaxy rest-frame luminosity to $z=0$, we find
  that galaxies are underluminous for a given \mbh, even neglecting growth
  by accretion.
\item The passively evolved correlations are well described by a
  relationship of the form $\Delta \log$\mbh $= \gamma \log (1+z)$
  with $\gamma=0.6\pm0.1$ and $\gamma=0.8\pm0.2$, respectively at
  fixed bulge and total host luminosity, taking into account selection
  effects.
\end{enumerate}

Considering that stellar populations must fade as they get older, and
considering that similar results have been found when studying the
correlations between \mbh\ and host galaxy velocity dispersion
\citep{TMB04,Woo++06,Woo++08} and stellar mass
\citep{Jah++09,Bennert11,SS13}, we are forced to conclude that the
co-evolution of galaxies and black holes is non-trivial, in the sense
that systems do not stay on the correlation as they evolve. At least
for active galaxies in the range of black hole and stellar masses that
can be analyzed with current technology, it appears that the growth of
the black hole predates that of the bulge \citep{Cro06}. However,
given the complexity and variety of processes involved, direct
comparisons with detailed numerical simulations are needed to further
our understanding of the co-evolution of black holes and their hosts.
Recent cosmological simulations including some prescriptions for black
hole growth and feedback have been shown to reproduce the observations
at least at $z<1$ \citep{DeG++15}. It will be interesting to carry out
similar detailed comparisons, taking into account errors and
observational selection functions, for a variety of models
\citep[e.g.][]{SiJ++15,T+K16,Vol++16} and extending to higher
redshifts. These comparisons will provide a powerful test of the
various recipes that have been adopted to describe accretion and star
formation physics at sub-grid level in numerical simulations.

Looking at the future, the sample of lensed quasars that can be
analyzed with high fidelity is going to grow. Currently, ultra deep
\hst\ imaging data have been obtained for six additional strongly
lensed systems\footnote{WFI2033$-$4723 and HE1104$-$1805 from H0LiCOW
  program \textit{HST}-GO-12889 (PI: Suyu); SDSS1206$+$4332,
  HE0047$-$1756, SDSS0246$-$0825 and HS2209$+$1914 as part of Program
  \textit{HST}-GO-14254 (PI: Treu).} and their analysis will be
described in a forthcoming paper. The sample of lensed quasars and
their hosts that can be studied at high fidelity is likely to continue
to grow as more such systems are discovered in wide field imaging and
spectroscopic surveys \citep[e.g.][]{Agn++15,Mor++16,Sch++16,Ost++17}.

\section*{Acknowledgements}
Based in part on observations made with the NASA/ESA Hubble Space
Telescope, obtained at the Space Telescope Science Institute, which is
operated by the Association of Universities for Research in Astronomy,
Inc., under NASA contract NAS 5-26555. These observations are
associated with programs \# 9744, 12889, 14254. Financial support was
provided by NASA through grants from the Space Telescope Science
Institute.

We are grateful to Vivien Bonvin, Geoff C.-F. Chen, Frederic Courbin, Matthew A. Malkan,
Cristian E. Rusu, Jong-Hak Woo, and Andreas Schulze for useful comments and suggestions
that improved this manuscript. We thank Chien Peng for his help with
the estimates of black hole mass.
X.D. is supported by the China Scholarship Council. T.T. acknowledges
support by the Packard Foundations through a Packard Research
Fellowship and by the NSF through grants AST-1450141 and AST-1412315.
S.H.S. gratefully acknowledges support from the Max Planck Society
through the Max Planck Research Group.  C.E.R. acknowledges support
from the NSF grant AST-1312329.  D.S. acknowledges funding support
from a {\it {Back to Belgium}} grant from the Belgian Federal Science
Policy (BELSPO).  K.C.W. and D.P. is supported by an EACOA Fellowship
awarded by the East Asia Core Observatories Association, which
consists of the Academia Sinica Institute of Astronomy and
Astrophysics, the National Astronomical Observatory of Japan, the
National Astronomical Observatories of the Chinese Academy of
Sciences, and the Korea Astronomy and Space Science
Institute. V.N.B. gratefully acknowledges assistance from a National
Science Foundation (NSF) Research at Undergraduate Institutions (RUI)
grant AST-1312296.  Note that findings and conclusions do not
necessarily represent views of the NSF.  V.B. acknowledge the support
of the Swiss National Science Foundation (SNSF). 




\input{paper_sim_II.bbl}



\newpage
\begin{table}
\setcounter{table}{0}
\centering
    \caption{Properties of AGNs in the distant sample.}\label{result}
     \resizebox{9cm}{!}{
     \begin{tabular}{ c c c c c}
     \hline
     Object & Line(s) Used& redshift& log \mbh & log \lhost$_{,R}$ \\
     &&&$(M_{\odot}$)&$(L_{\odot}$)\\
     &&&$\pm 0.4$ dex&$\pm 0.2$ dex\\
     \hline\hline
     HE~0435 & \Mgii& 1.693 & 8.61& 10.96 \\
     RXJ~1131$_{\rm bulge}$ & \Mgii/\Hb&0.654 & 8.26/8.41& 10.58\\  
     RXJ~1131$_{\rm disk}$ & \Mgii/\Hb&0.654 & 8.26/8.41& 11.12\\   
\hline\hline
\multicolumn{4}{l}{Lensed AGNs from P06}
\\ \hline
RXJ~1131& \Hb& 0.66& 7.90& 11.02\\
SDSS~1226-0006& \Mgii& 1.12& 8.41& 10.74\\
FBQ~0951+2635& \Mgii& 1.24& 8.57& 10.25\\
CTQ~414& \Mgii/\Civ& 1.29& 7.78/8.19& 10.87\\
B~0712+472& \Mgii& 1.34& 7.44& 10.90\\
SBS~0909+532& \Mgii& 1.38& 9.13& 10.54\\
Q~0957+561& \Mgii/\Civ& 1.41& 9.06/8.97& 11.79\\
FBQ~1633+3134& \Mgii/\Civ& 1.52& 8.84/8.91& 11.08\\
SDSS~0924+0219& \Mgii& 1.54& 7.61& 11.09\\
B~1030+071& \Mgii& 1.54& 8.13& 11.06\\
SDSS~1335+0118& \Mgii& 1.57& 8.77& 10.97\\
B~1600+434& \Mgii& 1.59& 7.56& 10.98\\
HE~0047-1756& \Civ& 1.66& 8.83& 11.24\\
HE~0435& \Civ& 1.69& 8.36& 11.12\\
PG~1115+080& \Mgii/\Civ& 1.72& 8.67/8.63& 11.08\\
SBS~1520+530& \Civ& 1.86& 8.60& 10.82\\
HE~2149-2745& \Civ& 2.03& 9.48& 11.47\\
HE~1104-1805& \Civ& 2.32& 9.03& 11.48\\
Q~1017-207& \Civ& 2.55& 8.88& 11.69\\
H~1413+117& \Civ& 2.55& 8.08& 11.48\\
MG~0414+0534& \Hb& 2.64& 9.07& 11.41\\
J~1004+1229& \Civ& 2.65& 8.97& 11.70\\
Q~0142-110& \Civ& 2.72& 9.01& 11.37\\
LBQS~1009-0252& \Mgii/\Civ& 2.74& 8.51/8.70& 11.48\\
RXJ~0911+0551& \Civ& 2.80& 8.57& 10.93\\
PMNJ~1632-0033& \Civ& 3.42& 8.25& 11.35\\
B~1422+231& \Mgii/\Civ& 3.62& 8.93/9.34& 11.60\\
BRI~0952-0115& \Civ& 4.5& 8.80& 11.95\\
\hline\hline
\multicolumn{4}{l}{Non-lensed AGNs from P06}
\\ \hline
PKS~0440-00&\Mgii&0.844&8.09&11.29\\
MGC~2214+3550A& \Mgii& 0.88& 8.76& 11.13\\
MGC~2214+3550B& \Mgii& 0.88& 8.26& 10.59\\
3C~422&\Mgii&0.942&9.04&11.64\\
PKS~0938+18&\Mgii&0.943&8.53&11.16\\
SGP5:46&\Mgii&0.955&8.03&10.92\\
LBQS~1009-0252c& \Civ& 1.63& 8.88& 11.34\\
RXJ~0921+4528A& \Mgii/\Civ& 1.65& 9.09/8.71& 11.12\\
RXJ~0921+4528B& \Mgii/\Civ& 1.65& 8.82/8.4& 11.38\\
SGP4:39&\Civ&1.716&8.07&10.49\\
MZZ~11408&\Civ&1.735&8.00&10.82\\
SGP2:36&\Civ&1.756&8.87&11.26\\
MZZ~1558&\Civ&1.829&8.80&10.94\\
SGP2:25&\Civ&1.868&8.45&11.27\\
MZZ~4935&\Civ&1.876&8.02&10.43\\
SGP3:39&\Civ&1.964&8.69&11.38\\
SGP2:11&\Civ&1.976&8.69&11.04\\
4C~45.51&\Mgii/\Civ&1.992&8.60/8.48&12.18\\
MZZ~9592&\Civ&2.71&8.47&11.49\\
MZZ~9744&\Civ&2.735&8.52&11.09\\
\hline\hline
\multicolumn{4}{l}{Non-lensed AGNs from P15} & (\lbulge/ $L_{\rm total}$)
\\ \hline
S09$^a$& \Hb& 0.3545& 7.99& 11.03/11.05\\
S10& \Hb& 0.3513& 8.44& 10.60/11.03\\
S12& \Hb& 0.3583& 8.78& 10.11/11.02\\
S21& \Hb& 0.3546& 8.93& 10.38/11.30\\
S16& \Hb& 0.3702& 8.02& 9.89/10.53\\
S23& \Hb& 0.3513& 8.82& 10.27/11.04\\
S24& \Hb& 0.3619& 8.20& 10.95/11.14\\
S26& \Hb& 0.3692& 8.00& 10.47/10.76\\
S27& \Hb& 0.3669& 7.85& 11.00/11.07\\
S01& \Hb& 0.3594& 8.15& 10.41/10.87\\
S02& \Hb& 0.3544& 8.02& 10.43/10.68\\
S03& \Hb& 0.3584& 8.17& 9.64/11.00\\
S04& \Hb& 0.3579& 8.11& 10.51/10.98\\
S05& \Hb& 0.3535& 8.73& 9.75/10.94\\
S06& \Hb& 0.3688& 7.72& 9.53/10.95\\
S07& \Hb& 0.352& 8.53& 10.28/11.02\\
S08& \Hb& 0.3586& 7.74& 10.02/10.89\\
S11& \Hb& 0.3559& 7.76& 10.50/10.90\\
SS1& \Hb& 0.3566& 7.73& 10.13/10.94\\
SS2& \Hb& 0.3671& 7.56& 10.76/10.76\\
SS5& \Hb& 0.3735& 7.98& 10.02/10.66\\
     \hline
     \end{tabular}}
\end{table}
\begin{table}
\setcounter{table}{0}
\centering
    \caption{---continued.}
    \resizebox{9cm}{!}{
     \begin{tabular}{ c c c c c}
     \hline
\multicolumn{4}{l}{Non-lensed AGNs from P15} & log \lbulge/log $L_{\rm total}$\\
\hline\hline
S31& \Hb& 0.3506& 8.19& 10.76/11.00\\
SS6& \Hb& 0.3588& 7.39& 9.67/10.47\\
SS7& \Hb& 0.3613& 7.67& 10.08/10.74\\
SS8& \Hb& 0.3655& 7.85& 9.96/10.92\\
SS9& \Hb& 0.3702& 7.95& 10.82/10.82\\
SS10& \Hb& 0.3658& 8.10& 10.61/10.81\\
SS11& \Hb& 0.3731& 7.83& 10.04/10.83\\
SS12& \Hb& 0.3629& 8.15& 10.54/10.66\\
SS13& \Hb& 0.3743& 7.69& 10.60/10.60\\
S28& \Hb& 0.3678& 8.12& 10.70/10.91\\
SS14& \Hb& 0.3706& 7.45& 10.48/10.48\\
S29& \Hb& 0.3574& 7.95& 10.03/10.69\\
SS18& \Hb& 0.3585& 7.51& 9.81/10.62\\
W11& \Hb& 0.565& 7.95& 10.73/10.73\\
W22& \Hb& 0.5652& 8.68& 11.27/11.27\\
W12& \Hb& 0.5623& 8.94& 10.49/10.98\\
W20& \Hb& 0.5761& 8.60& 11.03/11.03\\
W16& \Hb& 0.578& 7.86& 10.81/10.81\\
W8& \Hb& 0.5712& 8.74& 11.01/11.01\\
W3& \Hb& 0.576& 8.76& 10.28/10.88\\
SS15& \Hb& 0.3593& 7.44& 10.36/10.36\\
W1& \Hb& 0.5736& 8.84& 10.71/10.95\\
W4& \Hb& 0.5766& 8.28& 11.01/11.01\\
W5& \Hb& 0.5767& 8.29& 10.96/10.96\\
SS3& \Hb& 0.3566& 7.51& 10.10/10.78\\
SS4& \Hb& 0.3629& 7.85& 10.81/10.81\\
W17& \Hb& 0.5617& 8.31& 10.12/10.80\\
W2& \Hb& 0.572& 9.07& 10.91/10.91\\
W10& \Hb& 0.5711& 7.94& 10.51/10.77\\
W14& \Hb& 0.5617& 8.68& 10.97/10.97\\
W9& \Hb& 0.5654& 8.70& 10.89/10.96\\	
J033252-275119& \Mgii& 1.227& 8.87& 10.53/11.28\\
J033243-274914& \Mgii& 1.900& 9.17& 11.73/11.73\\
J033239-274601& \Mgii& 1.220& 8.24& 11.22/11.22\\
J033226-274035& \Mgii& 1.031& 7.85& 10.37/11.46\\
J033225-274218& \Mgii& 1.617& 8.08& 11.48/11.48\\
J033210-274414& \Mgii& 1.615& 8.30& 11.52/11.52\\
J033200-274319& \Mgii& 1.037& 7.75& 10.47/10.47\\
J033229-274529& \Mgii& 1.218& 8.37& 11.39/11.39\\
J123553+621037& \Mgii& 1.371& 8.27& 10.73/11.66\\
J123618+621115& \Mgii& 1.021& 8.35& 10.11/11.37\\
J123618+621115& \Mgii& 1.450& 8.77& 11.45/11.45\\
158$^b$& \Mgii& 0.717& 7.28& 10.71/10.89\\		
170& \Mgii& 1.065& 7.07& 10.16/10.46\\		
271& \Mgii& 0.960& 7.43& 10.46/11.36\\		
273& \Mgii& 0.970& 8.23& 10.19/10.45\\		
305& \Mgii& 0.544& 8.61& 11.02/11.14\\		
333& \Mgii& 1.044& 7.90& 10.45/10.91\\		
339& \Mgii& 0.675& 7.95& 10.83/11.00\\		
348& \Mgii& 0.569& 8.11& 10.64/11.17\\		
379& \Mgii& 0.737& 9.14& 11.03/11.70\\		
413& \Mgii& 0.664& 7.05& 10.19/10.74\\		
417& \Mgii& 0.837& 8.37& 10.18/10.87\\		
465& \Mgii& 0.740& 8.02& 10.67/11.47\\		
516& \Mgii& 0.733& 7.93& 11.02/11.38\\		
540& \Mgii& 0.622& 7.61& 10.84/11.30\\		
597& \Mgii& 1.034& 8.12& 10.80/10.91\\		
712& \Mgii& 0.841& 8.65& 11.22/11.59\\		
     \hline
     \end{tabular}}
     \begin{tablenotes}
      \small
      \item Note:$-$ Column 1: object ID. Column 2: Emission line used to estimate \mbh .
      Column 3: redshift as listed in the literature. Column 4: \mbh\ calibrated from Eq.~\ref{recipe}
      using the corresponding lines. 
      Column 5: Inferred rest-frame R-band luminosity not corrected for evolution. Note that all
      the host galaxies in P06 are assumed to be pure ellipticals.\\
      $^a$ ID taken from \citet{Park15}.\\
      $^b$ ID taken from \citet{SS13}.
           \end{tablenotes}
\end{table}

\begin{table}
\setcounter{table}{1}
\centering
    \caption{Properties of local AGNs.}\label{local}
     \begin{tabular}{ c c c c}
     \hline
     Object & redshift& log \mbh& log \lbulge$_{,R}$/log $L_{\rm total}$$_{,R}$ \\
     &&$(M_{\odot}$)&$(L_{\odot}$)\\
     &&&$\pm 0.2$ dex\\
     \hline\hline
     
3C120& 0.03301& 7.71$\pm$0.21& 10.51/10.51\\
3C390.3& 0.0561& 8.43$\pm$0.10& 10.55/10.55\\
Ark120& 0.03271& 8.15$\pm$0.06& 10.23/10.69\\
Mrk79& 0.02219& 7.69$\pm$0.12& 9.62/10.01\\
Mrk110& 0.03529& 7.37$\pm$0.11& 9.42/9.93\\
Mrk279& 0.03045& 7.51$\pm$0.11& 9.87/10.32\\
Mrk335& 0.02579& 7.12$\pm$0.11& 9.61/10.00\\
Mrk590& 0.02639& 7.65$\pm$0.07& 9.98/10.45\\
Mrk817& 0.03146& 7.66$\pm$0.07& 9.34/10.32\\
PG0052+251& 0.155& 8.54$\pm$0.09& 11.34/11.34\\
PG0804+761& 0.1& 8.81$\pm$0.05& 10.81/10.81\\
PG0844+349& 0.064& 7.94$\pm$0.18& 10.54/10.54\\
PG1211+143& 0.0809& 8.13$\pm$0.13& 10.43/10.43\\
PG1226+023& 0.15834& 8.92$\pm$0.09& 11.64/11.64\\
PG1229+204& 0.06301& 7.83$\pm$0.21& 10.30/10.70\\
PG1411+442& 0.0896& 8.62$\pm$0.14& 10.58/10.58\\
PG1613+658& 0.129& 8.42$\pm$0.20& 11.47/11.47\\
PG1700+518& 0.292& 8.86$\pm$0.10& 11.51/11.51\\
PG2130+099& 0.06298& 7.55$\pm$0.17& 9.82/10.42\\
     \hline
     \end{tabular}
     \begin{tablenotes}
      \small
      \item Note:$-$ Local AGNs measurements, taken from \citet{Ben++10}. 
      Following \citet {Park15}, we adopted virial factor as $\log f=0.71$. Note that
      \citet{Ben++10} adopted $\log f=0.74$.
     \end{tablenotes}
\end{table}

\appendix

\section{The K-correction for the RXJ1131 host}
\label{App1}
\begin{figure}
\centering
\subfloat[
K-correction from observed F814w band to the rest-frame R-band magnitude, as a function of stellar population age.
The colors are calculated based GALAXEV stellar evolution track with metallicities of  $Z=0.4Z_{\odot}$, $1.0Z_{\odot}$,
$2.5Z_{\odot}$ for E-type and Sb-type galaxies.
]
{\includegraphics[width=0.36\textwidth]{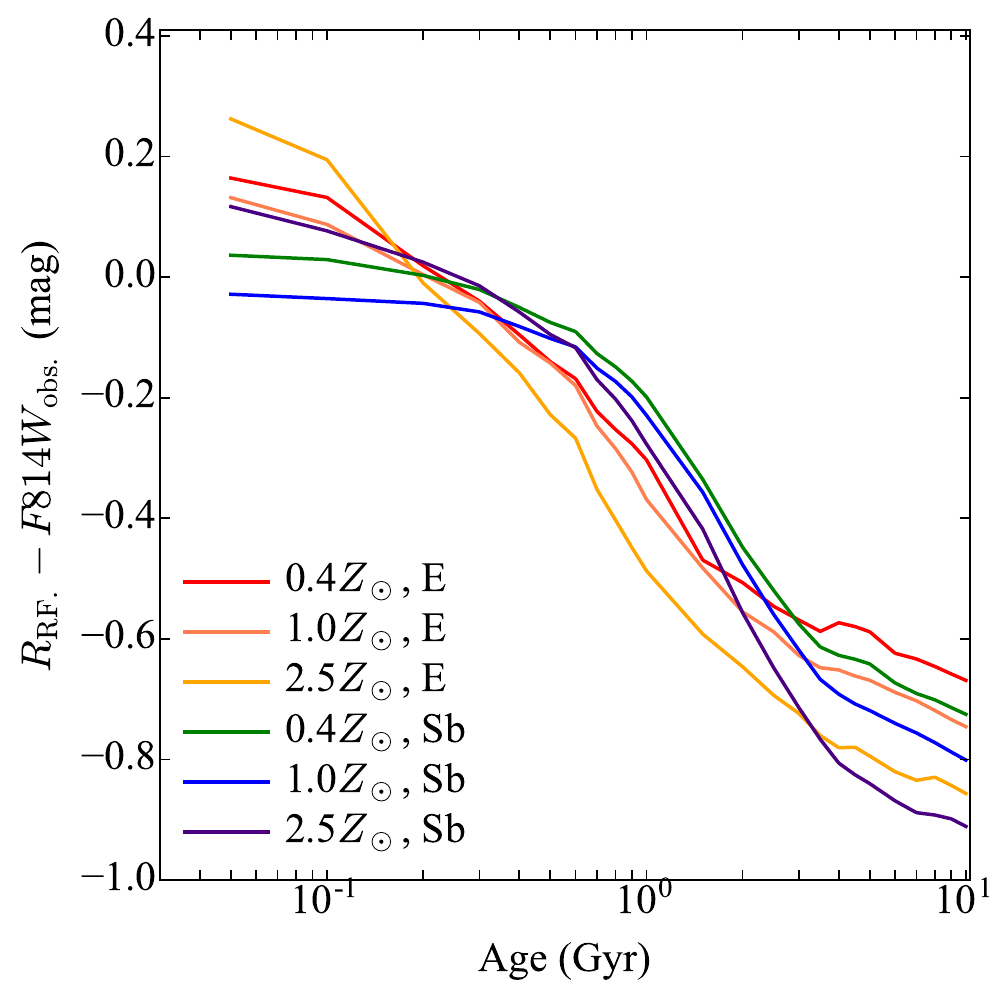}}\\
\subfloat[
Color map (the rest-frame R magnitude - observed F814W magnitude) of RXJ1131,
calculated by SED fitting with three broadband imaging (F555W, F814W and F160W). 
For bulge region, a direct measurement of SED is affected by the residual AGN contamination, and hence half blue and half red.
For the disk region, i.e. around the area where lensing-distorted
spiral arm patterns and star forming regions are clearly
visible, $\Delta$mag is approximately $-0.3$.
]
{\includegraphics[width=0.4\textwidth]{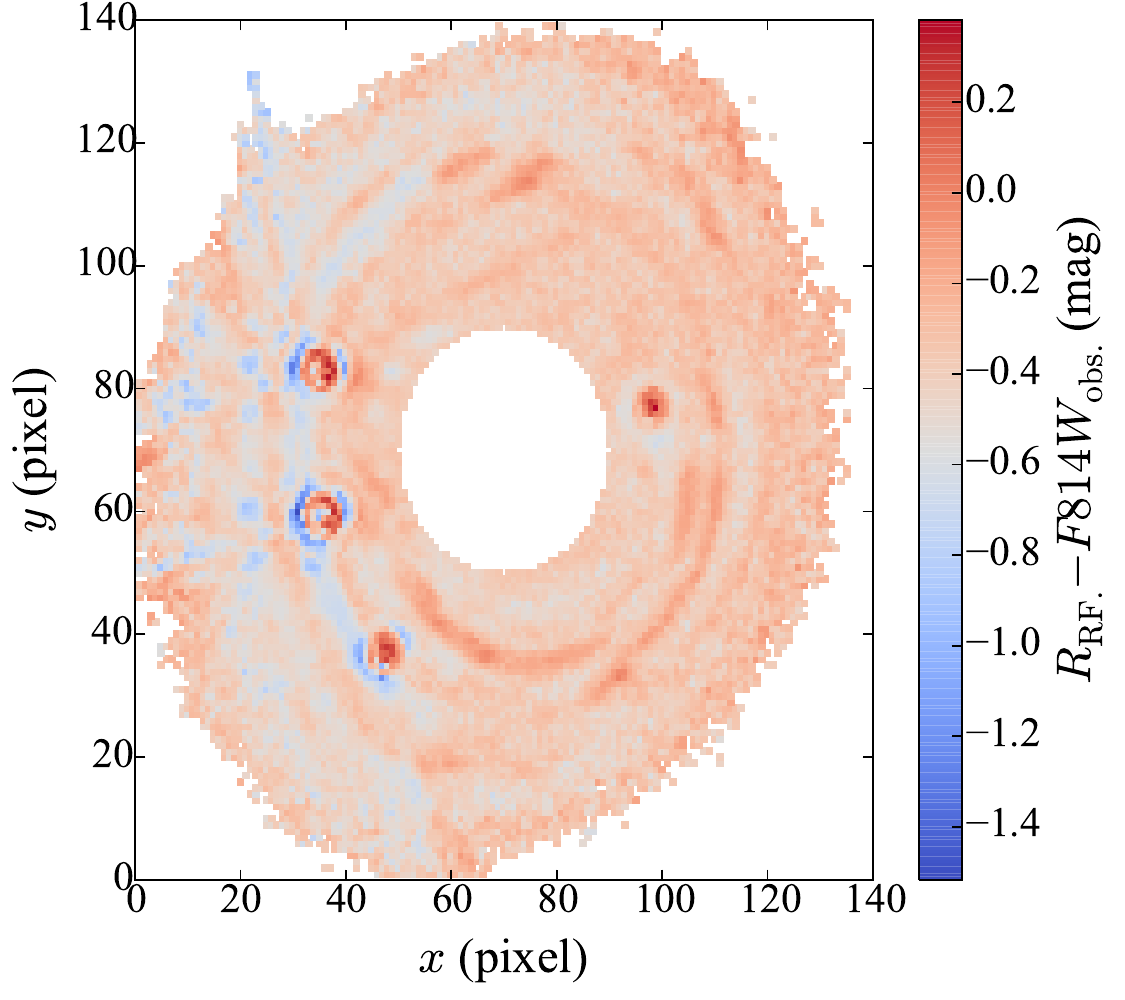}}
\caption{\label{fig:1131_sed} Illustration of the K-correction for RXJ1131.}
\end{figure}

We apply the K-correction to the observed magnitudes to obtain the rest-frame R-band magnitude.
At the redshift of RXJ1131 ($z_s=0.654$), the conversion from F814W to R-band magnitude depends on the
adopted SED templates, as shown in Fig.~\ref{fig:1131_sed}, panel(a).
Therefore, we determine directly the K-correction of the disk component through SED fitting with the multi-band images (F555W, F814W, F160W), available in the archive (GO-9744; PI: C.~S.~Kochanek).

Briefly, we fit the SED for each pixel by using {\sc Fast} \citep{Kriek2009} based on {\sc Galaxev} stellar evolution track
\citep{bruz2003}, assuming the solar metallicity, exponentially declining star formation history and
\citet{Calzetti2000} dust extinction law, while the redshift is fixed to the spectroscopic one.
The error for each pixel is calculated based on empty regions of images. We then derive the rest-frame R-band
magnitude by using the best-fit template for each pixel, and see the offset from the observed F814W magnitude.
As shown in Fig.~\ref{fig:1131_sed}, panel (b), $\Delta {\rm mag}_{\rm disk}$(R$-$F814W)$\approx-0.3$
is an appropriate estimate for the disk region. 
For the bulge, a direct measurement of SED is affected by the residual AGN contamination, and hence the
arc corresponding to the bulge is half blue and half red. Thus, we assume
that the age of stellar populations are $>3$ Gyr, where the correction hardly change (see Fig.~\ref{fig:1131_sed}, panel (a)),
and adopt $\Delta {\rm mag}_{\rm bulge}$(R$-$F814W)$\approx-0.7$. 

\bsp	
\label{lastpage}
\end{document}